\documentclass[a4paper,12pt]{article}
\usepackage{amsmath,amssymb,exscale}   
\usepackage{array,multicol}
\usepackage{afterpage,float,flafter}   
\usepackage{epsfig,rotating,pifont,fancybox}   
\usepackage{axodraw}
\input{axodraw.sty}
\makeatletter   
\@addtoreset{equation}{section}   
\makeatother   
   
\newcommand{\Id}{{\sf I}\hspace{-5.pt}{\rm 1}}

\newcommand\be{\begin{equation}}
\newcommand\ee{\end{equation}}
\newcommand{\brr}{\begin{eqnarray}}
\newcommand{\err}{\end{eqnarray}}

\newcommand{\bea}{\begin{eqnarray}}   
\newcommand{\eea}{\end{eqnarray}}

\def\simlt{\stackrel{<}{{}_\sim}}
\def\simgt{\stackrel{>}{{}_\sim}}

\title{   
\vspace*{-0.8cm}   
\begin{flushright}   
\normalsize{      
\texttt{hep-ph/0302189}}\\ 
\end{flushright}    
\normalsize
\vspace{1cm}
\Large
{\bf 
New Ideas in Symmetry Breaking~\footnote{Based on lectures given
at the 2002 Theoretical Advanced Study Institute (TASI-02), held at the
University of Colorado, CO, USA, from June 3-28, 2002, and at the Summer
Institute 2002 (SI2002) held at Fuji Yoshida, Japan, from August 13-20,
2002.}}
\vspace*{.5cm} \author{\large {\bf 
M.~Quir{\'o}s}\\ \\
\emph{Instituci\'o Catalana de Recerca i Estudis Avan\c{c}ats (ICREA)}\\
\emph{Theoretical Physics Group, IFAE}\\ 
\emph{E-08193 Bellaterra (Barcelona) Spain}}}
\date{}

\begin{document}
\maketitle
\thispagestyle{empty}
\vspace*{.5cm}

\begin{abstract}\noindent
Some old and new ideas on symmetry breaking,
based on the presence of extra dimensions that have been the subject
of a very fast development and intensive studies during the last
years, will be presented in these lectures. Special attention will be
devoted to the various compactification mechanisms, including toroidal
and orbifold compactifications, and to non-trivial boundary conditions
or Scherk-Schwarz compactification. Also symmetry breaking by Wilson lines, or
Hosotani breaking characteristic of non-simply connected compact
manifolds will be analyzed in some detail. The different mechanisms
will be applied to the breaking of the most relevant symmetries in
particle physics: supersymmetry and gauge symmetry. The required
background for these lectures is Quantum Field Theory, Supersymmetry
and some rudiments of Kaluza-Klein theory. The different sections will
be illustrated with examples.
\end{abstract}
\vspace{3.5cm}   
   
\begin{flushleft}   
February 2003 \\   
\end{flushleft}
\newpage

\section{Introduction}

Symmetry breaking is one of the main issues in contemporary particle
physics. Its implementation in a perturbative quantum field theory has
led to the notion of spontaneous symmetry breaking with the presence
of a massless Goldstone particle for the case of a global symmetry
(Goldstone theorem~\cite{Goldstone}) and a massive Higgs particle for
a local symmetry (Higgs mechanism~\cite{Higgs}). These ideas based on
four dimensional field theories are nowadays thoroughly explained in
many textbooks in quantum field theory~\cite{Peskin}.

The idea of unification of all interactions led to the introduction of
extra dimensions starting from the pioneer work of Kaluza and Klein
who attempted to unify gravity and electromagnetism through the
presence of a fifth dimension~\cite{KK}. More recently the attempts to
unify gravity with electroweak and strong interactions in a consistent
quantum theory have led to the modern string theories that incorporate
supersymmetries and are described in ten (heterotic, type II and type
I strings) or eleven (M-theory) space-time
dimensions~\cite{strings}. The recent string dualities relating all
string constructions~\cite{dualities} as well as the presence in the
spectrum of superstrings and supergravities of branes embedded in the
higher dimensional space (as e.~g. in type I strings Dp-branes with
their world-sheet spanning a $p+1$-dimensional space-time) opened the
possibility that the non-gravitational sector could live in a $p+1$
dimensional hypersurface~\footnote{The gravitational sector must
propagate in the whole higher-dimensional space time.}. Moreover the
size of both the string scale and the compactification radius can be
lowered from the Planck scale to the TeV
range~\cite{stringTeV,radiusTeV} thus making contact with the
phenomenology of present and future colliders~\cite{pheno}. Therefore
if the Standard Model lives in a $Dp$-brane with $p>3$ this means
that the Standard Model fields feel extra dimensions in its space-time
propagation. This in turn opened a plethora of new possibilities for
symmetry breaking associated with the different compactifications that
extra dimensions can experience.

These possibilities will be described in these lectures in some
detail. Some of them are based on the possible compactification of
extra dimensions. This compactification can break the
higher-dimensional Lorentz invariance (as in toroidal
compactifications) or the higher-dimensional Poincare invariance as in
orbifolds where translational invariance is explicitly broken and
four-dimensional fixed points can appear where localized (or twisted)
states can propagate~\cite{orbifolds}. The compactification of extra
dimensions can also introduce non-trivial boundary conditions, a
mechanism known as Scherk-Schwarz breaking~\cite{SSbreaking}. Finally the extra
dimensional components of gauge fields can acquire a constant
background or vacuum expectation value and a symmetry can then be
broken by a Wilson line, a mechanism known as Hosotani
mechanism~\cite{Hosotani}.

These three lectures will be organized in the following way. A general
overview of all these mechanisms will be given in section~\ref{extra}
without explicit mention to the particular symmetry that is
broken. Sections~\ref{susy} and \ref{gauge} will be devoted to the
particularly interesting cases in particle physics where the symmetry
is identified with supersymmetry and gauge symmetry,
respectively. Since this is not an introduction on these general
topics the reader is supposed to have a knowledge on supersymmetric
and (non-abelian) gauge field theories as well as some notions on
Kaluza-Klein theories. As we said above the required background on
quantum field theories is provided by standard textbooks~\cite{Peskin}
while an introduction to supersymmetric theories can be found in
Ref.~\cite{susy}. As for Kaluza-Klein theories an introduction has
been provided in these Tasi lectures~\cite{TasiKK}. The detailed
contents of these lectures goes as follows.


\vspace{.5cm}

\begin{center}
\underline{{\sc Table of contents}}
\end{center}

\vspace{1cm}

\noindent{\sc LECTURE I: EXTRA DIMENSIONS AND SYMMETRY BREAKING}

\begin{itemize}
\item
Compactification
\item
Scherk-Schwarz mechanism
\item
Orbifolds
\item
Scherk-Schwarz breaking in orbifolds
\item
Scherk-Schwarz as Hosotani breaking
\end{itemize}

\vspace{.5cm}
\noindent{\sc LECTURE II: SUPERSYMMETRY BREAKING}

\begin{itemize}
\item
Supersymmetry breaking by orbifolding

\begin{itemize}
\item
Vector multiplets
\item
Hypermultiplets
\end{itemize}

\item
Supersymmetry breaking by Scherk-Schwarz compactification

\begin{itemize}
\item
Bulk breaking
\item
Brane breaking
\end{itemize}

\item
Supersymmetry breaking by Hosotani mechanism

\begin{itemize}
\item
Super-Higgs effect

\item
Radiative determination of the Scherk-Schwarz parameter

\item
Brane assisted Scherk-Schwarz supersymmetry breaking
\end{itemize}
\end{itemize}

\vspace{.5cm}
\noindent{\sc LECTURE III: GAUGE SYMMETRY BREAKING}

\begin{itemize}
\item
Gauge breaking by orbifolding
 
\begin{itemize}
\item
Rank preserving
\item
Rank lowering
\end{itemize}
\item
Gauge breaking by the Hosotani mechanism
\item
Top assisted electroweak breaking
\end{itemize}


\section{Extra dimensions and symmetry breaking}

\label{extra}
In this lecture we will review some general ideas dealing with
symmetry breaking that can be realized in theories with extra
dimensions. We will start by defining the compactification mechanisms
on smooth manifolds (torii) with trivial and with twisted boundary
conditions. The former being known as ordinary and the latter as
Scherk-Schwarz compatification~\cite{SSbreaking}. Then we will
consider compactification on singular manifolds (orbifolds), with
singularities concentrated on the fixed points~\cite{orbifolds}. In
particular we will study the compatibility of orbifolds and
Scherk-Schwarz compatification. Finally we will interpret the
Scherk-Schwarz breaking as a Hosotani breaking~\cite{Hosotani} where
the extra dimensional component of a gauge boson acquires a vacuum
expectation value (VEV). In the present lecture we will only review
general ideas on symmetry breaking by orbifold and/or Scherk-Schwarz
compatification. We will postpone the consideration of specific
symmetries (as supersymmetry of gauge symmetry) to the second and
third lectures.

\subsection{Compactification}\label{compactification}
We will consider a $D$-dimensional theory ($D=4+d$) with $d$ extra
dimensions and an action defined as
\begin{equation}\label{sd}
S_D=\int d^Dz\ \mathcal{L}_D[\phi(z)] .
\end{equation}
We say that the theory is compactified on $\mathcal{M}_4\times C$,
where $\mathcal{M}_4$ is the Minkowski space-time and $C$ a compact
space if the coordinates of the $D$-dimensional space can be split as
$z^M=(x^\mu,y^m)$, ($\mu=0,1,2,3$; $m=1,\dots,d$) and the coordinates
$y^m$ describe the compact space $C$. The four dimensional (4D)
Lagrangian is obtained after integration of the compact coordinates
$y^m$ as
\begin{equation}\label{4Dlag}
\mathcal{L}_4=\int d^dy\ \mathcal{L}_D[\phi(x^\mu,y^m)]\ .
\end{equation}
The Lagrangian (\ref{4Dlag}) contains propagation and interaction of
massless and massive fields. For energies $E\ll \ell_C^{-1}$, where
$\ell_C$ is the typical size of $C$, heavy fields can be integrated
out and the Lagrangian (\ref{4Dlag}) describes an effective four
dimensional theory of massless fields with non-renormalizable
(higher dimensional) operators. 

In general we can write $C=M/G$, where $M$ is a (non-compact) manifold
and $G$ is a discrete group acting {\it freely} on $M$ by operators
$\tau_g:\ M\to M$ for $g\in G$. $M$ is the $covering$ space of
$C$. That $G$ is acting freely on $M$ means that only $\tau_{\imath}$
has fixed points in $M$, where $\imath$ is the identity in
$G$~\footnote{Trivially $\tau_{\imath} (y)=y,\ \forall y\in M$.}. The
operators $\tau_g$ constitute a representation of $G$, which means
that $\tau_{g_1 g_2}=\tau_{g_1}\cdot\tau_{g_2}$. $C$ is then
constructed by the identification of points $y$ and $\tau_g(y)$ that
belong to the same orbit
\begin{equation}\label{ident1}
y\equiv \tau_g(y)\ .
\end{equation}

After the identification (\ref{ident1}) physics should not depend on
individual points in $M$ but only on orbits (points in $C$) as
\begin{equation}\label{identlag}
\mathcal{L}_D[\phi(x,y)]=\mathcal{L}_D[\phi(x,\tau_g(y))] \ .
\end{equation}
A sufficient condition to fulfill Eq.~(\ref{identlag}) is
\begin{equation}\label{ordcomp}
\phi(x,\tau_g(y))=\phi(x,y)
\end{equation}
which is known as {\it ordinary} compactification. However condition
(\ref{ordcomp}), if sufficient is normally not necessary. In fact a
necessary and sufficient condition to fulfill Eq.~(\ref{identlag}) is
provided by
\begin{equation}\label{SScomp}
\phi(x,\tau_g(y))=T_g\phi(x,y)
\end{equation}
where $T_g$ are the elements of a global symmetry group of the
theory. Condition (\ref{SScomp}) is known as {\it Scherk-Schwarz}
compactification and will be the subject of the next section.

\subsection{Scherk-Schwarz mechanism}\label{SS}

As we described in the previous section the Scherk-Schwarz
compactification mechanism occurs when some {\it twist} transformation
corresponding to $g\in G$, $T_g$, is different from the identity. The
operators $T_g$ are a representation of the group $G$ acting on field
space, i.~e. they satisfy the property: $T_{g_1 g_2}=T_{g_1} T_{g_2},\
g_1,g_2\in G$. The latter property is easily proven by using the group
representation properties of operators $\tau_g$ acting on the space
$M$.

A few comments are in order now:

\begin{itemize}
\item
The ordinary compactification corresponds to $T_g=1,\ \forall g\in G$.

\item
Scherk-Schwarz compactification corresponds to $T_g\neq 1$ for some $g\in G$.

\item
For ordinary and Scherk-Schwarz compactifications fields are functions
on the covering space $M$.

\item
For ordinary compactification fields are also functions on the compact
space $C$.

\item
For Scherk-Schwarz compactification twisted fields are {\it not}
single-values functions on $C$.

\end{itemize}

\vspace{.5cm}
\begin{center}
\fbox{\sc Example}
\end{center}

\noindent
We will consider the simplest example of a compact space, the
circle. We will take one extra dimension $D=5$ (i.~e. $d=1$),
$M=\mathbb{R}$ (the set of real numbers), $G=\mathbb{Z}$ (the set of
integer numbers), and $C=S^1$ (the circle). The $n$-th element of the
group $\mathbb{Z}$ is represented by $\tau_n$ with
\begin{equation}\label{taun}
\tau_n(y)=y+2\pi n R,\ y\in \mathbb{R}
\end{equation}
where $R$ is the radius of the circle $S^1$. The identification
$y\equiv\tau_n(y)$ leads to the fundamental domain of length $2\pi R$
(the circle) as: $[y,y+2\pi R)$ or $(y,y+2\pi R]$. The interval must
be opened at one end because $y$ and $y+2\pi R$ describe the same
point in $S^1$ and should not be counted twice. Any choice for $y$
leads to an equivalent fundamental domain in the covering space
$\mathbb R$. A convenient choice is $y=-\pi R$ which leads to the
fundamental domain
\begin{equation}\label{domain}
(-\pi R,\pi R]
\end{equation}

The group $\mathbb{Z}$ has infinitely many elements but all of them
can be obtained from just one generator, the translation $2\pi
R$. Then only one independent twist can exist acting on the fields, as
\begin{equation}\label{twistT}
\phi(x,y+2\pi R)=T\, \phi(x,y)
\end{equation}
while twists corresponding to the other elements of $\mathbb{Z}$ are
just given by $T_n=T^n$.  

As we said above, $T$ must be an operator corresponding to a symmetry
of the Lagrangian. As we will see later on in these lectures this
symmetry can be a global, or local $SU(2)_R$, when the theory is
supersymmetric. Another candidate is a $\mathbb{Z}_2$ symmetry
associated to the invariance of the Lagrangian under the inversion
$\phi\to -\phi$~\footnote{This symmetry is present in the fermionic
sector and is associated to fermion number. It is used in field theory
at finite temperature~\cite{finiteT} where Euclidean time is
compactified on $S^1$.}. The case of $SU(2)_R$ will be analyzed in the
context of supersymmetric theories. Here we will analyze the simplest
$\mathbb{Z}_2$ case. There are fields with untwisted $T=1$ (bosons)
and fields with twisted $T=-1$ (fermions) boundary
conditions. Untwisted fields can be described by real functions on
$S^1$ while twisted fields are not single-valued functions on the
circle. Of course one can define single-valued functions on the
interval $[-\pi R,\pi R]$ by the definition $\phi(-\pi R)\equiv
T\phi(\pi R)=-\phi(\pi R)$ although still $\phi$ is not a
single-valued function on the circle $S^1$.


The case we have just studied can be easily generalized to that of
$p$-extra dimensions, $M=\mathbb{R}^p$, where $G=\mathbb{Z}^p$ and
$C=T^p$, the $p$-torus. In that case the torus periodicity is defined
by a lattice vector $\vec{v}=(v_1,\dots,v_p)$, where $v_i=2\pi R_i$
and $R_i$ are the different radii of $T^p$.  Twisted boundary
conditions are defined by $p$ independent twists given by $T_i$
\begin{equation}\label{twistgen}
\phi(x,y^j+ \delta_{ji} v^j)=T_{i}\,\phi(x,\vec{y})
\end{equation}
where $n_j=\delta_{ji}$ is the unitary vector along the $i$-th dimension.

\subsection{Orbifolds}\label{orbifolds}

Orbifolding is a technique normally used to obtain chiral fermions
from a (higher dimensional) vector-like
theory~\cite{orbifolds}. Orbifold compactification can be defined in a
similar way to ordinary or Scherk-Schwarz compactification. Let $C$ be
a compact manifold and $H$ a discrete group represented by operators
$\zeta_h:\ C\to C$ for $h\in H$ acting {\it non freely} on $C$. We mod
out $C$ by $H$ by identifying points in $C$ which differ by $\zeta_h$
for some $h\in H$ and require that fields defined at these two points
differ by some transformation $Z_h$, a {\it global or local symmetry}
of the theory,
\begin{eqnarray}\label{orbcomp}
y&\equiv & \zeta_h(y)\nonumber\\
\phi(x,\zeta_h(y))&= & Z_h \phi(x,y)\ .
\end{eqnarray}
The fact that $H$ acts non-freely on $C$ means that some
transformations $\zeta_h$ have fixed points in $C$. The resulting
space $O=C/H$ is not a smooth manifold but it has singularities at the
fixed points: it is called an orbifold.

\vspace{1cm}
\begin{center}
\fbox{\sc Example}
\end{center}

\vspace{1cm}

\noindent
We will continue with the simple example of the previous section with
$d=1$ and $C=S^1$. We can take in this case $H=\mathbb{Z}_2$ and the
orbifold is now $O=S^1/\mathbb{Z}_2$. The action of the only non-trivial
element of $\mathbb{Z}_2$ (the inversion) is represented by $\zeta$ where 
\begin{equation}\label{orbz2}
\zeta(y)=-y
\end{equation}
that obviously satisfies the condition
$\zeta^2(y)=\zeta(-y)=y\Rightarrow\zeta^2=1$. For fields we can write
as in (\ref{orbcomp})
\begin{equation}\label{z2field}
\phi(x,-y)=Z\,\phi(x,y)
\end{equation}
where using (\ref{orbz2}) and (\ref{z2field}) one can easily prove
that $Z^2=1$. This means that in field space $Z$ is a matrix that can
be diagonalized with eigenvalues $\pm 1$. The orbifold
$S^1/\mathbb{Z}_2$ is a manifold with boundaries: the fixed points are
co-dimension one boundaries. Not all orbifolds possess boundaries: for
example in $d=2$, $T^2/\mathbb{Z}_2$ is a ``pillow'' with four fixed
points without boundaries. A detailed description of six-dimensional
orbifolds can be found in Ref.~\cite{6Dorbi}.

\subsection{Scherk-Schwarz in Orbifolds}\label{SSorb}

In this section we will consider the case of Scherk-Schwarz
compactification in orbifolds. Remember that the starting point was a
non-compact space $M$ with a discrete group $G$ acting freely (without
fixed points) on the covering space $M$ by operators $\tau_g,\, g\in
G$ and defining the compact space $C=M/G$. The elements $g\in G$ are
represented on field space by operators $T_g$,
Eq.~(\ref{SScomp}). Subsequently we introduced another discrete group
$H$ acting non-freely (with fixed points) on $C$ by operators
$\zeta_h,\, h\in H$ and represented on field space by operators $Z_h$,
Eq.~(\ref{orbcomp}). We can always consider the group $H$ as acting on
elements $y\in M$ and then considering both $G$ and $H$ as subgroups
of a larger discrete group $J$. Then in general $\tau_g\cdot
\zeta_h(y)\neq\zeta_h\cdot\tau_g(y)$ which means that $g\cdot h\neq
h\cdot g$ and $J$ is {\it not} the direct product $G\otimes
H$. Furthermore the twists $T_g$ have to satisfy some consistency
conditions. In fact from Eqs.~(\ref{SScomp}) and (\ref{orbcomp}) one
can easily deduce a set of identities as

\begin{eqnarray}\label{comp}
T_g Z_h\,\phi(x,y)&=&\phi(x,\tau_g\cdot\zeta_h(y))\equiv
Z_{gh}\,\phi(x,y) \nonumber\\
Z_hT_g\,\phi(x,y)&=&\phi(x,\zeta_h\cdot\tau_g(y))\equiv
Z_{hg}\,\phi(x,y)\nonumber\\ 
T_{g_1}\, Z_h\,
T_{g_2}\,\phi(x,y)&=&\phi(x,\tau_{g_1}\cdot\zeta_h\cdot\tau_{g_2}(y))\equiv
Z_{g_1hg_2}\,\phi(x,y)
\end{eqnarray}

\noindent
where $g_1,g_2,h$ are considered as elements in the larger group
$J$. The conditions (\ref{comp}) impose compatibility constraints in
particular orbifold construction with twisted boundary conditions as
we will explicitly illustrate next.

\vspace{2cm}
\begin{center}
\fbox{\sc Example}
\end{center}

\noindent
We will continue here by analyzing the simple case of the orbifold
$S^1/\mathbb{Z}_2$ with twisted boundary conditions. In this case
there is only one independent group element for $G=\mathbb{Z}$ which
is the translation $\tau(y)=y+2\pi R$ while the orbifold group
$H=\mathbb{Z}_2$ contains only the inversion $\zeta(y)=-y$.

First of all, notice that the translation and inversion do not commute
to each other. In fact $\zeta\cdot\tau(y)=-y-2\pi R$ while
$\tau\cdot\zeta(y)=-y+2\pi R$. It follows then that
$\zeta\cdot\tau\cdot\zeta=\tau^{-1}$ and the group $J$ is the
semi-direct product $\mathbb{Z}\ltimes
\mathbb{Z}_2$~\cite{HallWilson}. Second, one can easily see that
$\tau\cdot\zeta\cdot\tau=\zeta$ which implies the consistency
condition on the possible twist operators
\begin{equation}\label{consistencia}
T\,Z\,T=Z\Longleftrightarrow ZTZ=T^{-1}
\end{equation}  
as can be easily deduced from Eq.~(\ref{comp}).

Since we have seen that $Z^2=1$, its eigenvalues are $\pm 1$ and $Z$
can be written in the diagonal basis as
\begin{equation}\label{zeta}
Z=\left(
\begin{array}{cc}
\sigma^3\,{\Id}_p& 0\\
0 & \pm 1
\end{array}\right)
\end{equation}
where $\Id_p$ is the identity matrix in $p$-dimensional field
space. This means that in the subspace spanned by some fields $Z$ can
be given either by $Z=\sigma^3$ or $Z=\pm 1$.

On the other hand being $T$ an operator corresponding to a global (or
local) symmetry of the theory it can be written as
\begin{equation}\label{te}
T=e^{2\pi i \vec\beta\cdot\vec\lambda}
\end{equation}
where $\lambda^a$ are the (hermitian) generators of the symmetry group
acting on field space. Using the consistency condition
(\ref{consistencia}) and (\ref{te}) leads to the condition
\begin{equation}\label{anticon}
\{\vec\beta\cdot\vec\lambda,Z \}=0
\end{equation}
for a generic $T$ that does not commute with $Z$. There is however a
singular solution for $[T,Z]=0$ which corresponds to $T=\pm 1$. Using
now (\ref{consistencia}) we can finally conclude that
\begin{eqnarray}\label{final}
Z&=&\sigma^3\Longrightarrow
T=e^{2\pi i (\beta_1\sigma^1+\beta_2\sigma^2)}\quad {\rm or} \quad
T=e^{2\pi i \frac{1}{2}\sigma^3}\\ Z&=&\pm
1\Longrightarrow T=\pm 1 \label{final2}
\end{eqnarray}
where the parameters $\beta_{1,2}$ are real-valued.  The previous
equations deserve some explanation. In the case of $Z=\sigma^3$ we are
implicitly considering the field subspace spanned by $SU(2)$ doublets
in which case $\vec\lambda=\vec\sigma$. This is typically the case
where the global symmetry of the Lagrangian is $SU(2)$. This case will
appear, as we will see next, in supersymmetric theories where gauginos
in vector superfields and scalar fields in hyperscalars transform as
doublets under an $SU(2)_R$ symmetry of the theory~\cite{PomQui}. In
that case, using the global residual invariance, we can rotate
$(\beta_1,\beta_2)\to(0,\omega)$ and consider twists given by
\begin{equation}\label{twistfin}
T=e^{2\pi i \omega\sigma^2}=\left(
\begin{array}{rr}
\cos 2\pi \omega & \sin 2\pi\omega\\
-\sin 2\pi\omega &\cos 2\pi \omega
\end{array}
\right)
\end{equation}
The twist (\ref{twistfin}) is a continuous function of $\omega$ and so
it is continuously connected with the identity that corresponds to the
trivial no-twist solution (i.~e. $\omega=0$). In this way
Eq.~(\ref{twistfin}) describes a {\it continuous} family of solutions
to the consistency condition (\ref{consistencia}). There is also a
{\it discrete} solution that is not connected with the identity and
corresponds to $\omega=1/2$ for $T=\exp(\pi i\sigma^3)=-\,\Id_2$ as
shown in Eq.~(\ref{final}). In this case $[Z,T]=0$.

In the case of $Z=\pm 1$ we are considering a discrete global symmetry
with even and odd fields under $y\to -y$. Now using
Eq.~(\ref{consistencia}) we obtain that boundary conditions can be
either periodic or anti-periodic, i.~e. $T=\pm 1$. For instance if the
global symmetry can be associated to fermion number, bosons (fermions)
have periodic (anti-periodic) boundary conditions. In particular this
is the case of field theory at finite
temperature~\cite{finiteT}. Another example is the case of
supersymmetric theories where we can use $R$-parity of $N=1$ four
dimensional supersymmetry as the global symmetry. Here ordinary
Standard Model fields are periodic while their supersymmetric partners
are antiperiodic~\cite{Barbieri}.

\subsection{Scherk-Schwarz as Hosotani breaking}\label{hosotani}

In theories compactified on a torus, or an orbifold, a symmetry can be
broken by two mechanisms that are not present in simply-connected
spaces: the Scherk-Schwarz and the Wilson/Hosotani mechanisms. As we
have seen above the Scherk-Schwarz mechanism is based on twists $T_g$
that represent the discrete group $G$ defining the torus in field
space by means of a global, or local symmetry of the Lagrangian. If
the symmetry is a {\it local} one the Scherk-Schwarz breaking is
equivalent to a Hosotani breaking, where an extra dimensional
component of the corresponding gauge field acquires a non-zero VEV.

For simplicity we will consider the case of a five-dimensional, $d=1$
theory compactified on the $S^1/\mathbb{Z}_2$ orbifold. In this case
the Scherk-Schwarz twist is $T=\exp(2\pi i \omega Q)$, i.~e.
\begin{equation}\label{twistW}
\phi(x,y+2\pi R)=e^{2\pi i \omega Q}\ \phi(x,y)
\end{equation}
where $Q$ corresponds to a given direction in the generator space and
$\omega$ is the corresponding parameter. A trivial solution to
Eq.~(\ref{twistW}) is
\begin{equation}\label{twistsol}
\phi(x,y)=e^{i\omega Q y/R}\ \widetilde{\phi}(x,y)
\end{equation}
where $\widetilde{\phi}(x,y+2\pi R)=\widetilde{\phi}(x,y)$ is a
periodic single-valued function that can be expanded in Fourier modes.
Obviously the symmetry generated by $Q$ is broken by the
five-dimensional kinetic term. 

Since the symmetry is a local one, there are associated gauge fields
$\vec{A}_M$ ($M=\mu,5$). If there is a VEV for $\vec{A}_5$ along the
$Q$-direction, $\langle \vec\lambda\cdot\vec A_5\rangle\equiv Q\langle
A_5^Q\rangle$ all non-singlet fields will receive a mass-shift
relative to their KK-values through their covariant derivatives
$\propto \langle A_5^Q\rangle$. In this representation all fields are
{\it periodic} (no twist). We can switch to the Scherk-Schwarz picture by
allowing for gauge transformations with {\it non-periodic}
parameters~\cite{GQWilson}. In particular the non-periodic gauge
transformation
\begin{equation}\label{gtrans}
U(y)=e^{2\pi i Q\langle A_5^Q\rangle y}
\end{equation}
transforms away the VEV and ends up with non-periodic fields with Scherk-Schwarz
parameter $\omega$ given by
\begin{equation}\label{HSS}
\langle A_5^Q\rangle=\frac{\omega}{R}
\end{equation}

\vspace{.5cm}
\begin{center}
\fbox{\sc Example}
\end{center}
\noindent
As an introduction to the next section we will consider here a
five-dimensional theory with $N=1$ global supersymmetry where there is
a global $SU(2)_R$ invariance acting on hyperscalars and
gauginos. Hyperscalars $\phi^i$ ($i=1,2$) are contained in
\fbox{hypermultiplets} $\mathbb{H}=(\phi^i,\psi)$ where $\psi$ is a
Dirac spinor and $\phi^i$ are complex scalars transforming as a
doublet under $SU(2)_R$~\cite{Sohnius}. Gauginos $\lambda^i$ ($i=1,2$)
are contained in \fbox{vector multiplets}
$\mathbb{V}=(A_M,\Sigma,\lambda^i)$ where $A_M$ are five-dimensional
gauge bosons and $\Sigma$ is a real scalar. $\lambda^i$ are
Symplectic-Majorana spinors transforming as doublets under $SU(2)_R$,
i.~e.~\cite{MirPes}
\begin{equation}\label{SympMaj}
\lambda^i=\left(
\begin{array}{c}
\lambda^i_L\\
\epsilon^{ij}\bar\lambda_{jL}
\end{array}
\right),\quad \bar\lambda_{jL}\equiv -i\sigma^2\left(\lambda^j_L\right)^{*}
\end{equation}
where $\lambda^i_L$ are Weyl fermions and $\epsilon^{ij}$ is the
two-dimensional antisymmetric tensor with $\epsilon^{12}=+1$ . To
summarize, the objects transforming under $SU(2)_R$ are the doublets
\begin{equation}
\left(
\begin{array}{c}
\phi^1\\
\phi^2
\end{array}
\right),\quad
\left(
\begin{array}{c}
\lambda^1\\
\lambda^2
\end{array}
\right)
\end{equation}
These fields are the only ones that can be given non-trivial boundary
conditions based on the $SU(2)_R$ global symmetry. This property will
be widely used in section~\ref{susy}.

Suppose now that $SU(2)_R$ is realized locally (see
section~\ref{susy}) and we define the orbifold breaking such that:
$(A_\mu^3,A_5^{1,2},\Sigma^{1,2})$ are even under the $\mathbb{Z}_2$
parity, while $(A_\mu^{1,2},A_5^3,\Sigma^3)$ are odd fields. This
amounts to defining the parities in the language of
section~\ref{SSorb} as
\begin{equation}\label{paridades}
Z_{A_5}=Z_{\Sigma}=-Z_{A_\mu}=\left(
\begin{array}{ll}
1& 0\\
0& \sigma^3
\end{array}\right)
\end{equation}
The five-dimensional kinetic terms for gauginos and hyperscalars can
be written as
\begin{equation}\label{cinetico}
\mathcal{L}_{kin}=\frac{i}{2}\bar\lambda \gamma^M \mathcal{D}_M\lambda
+\left|\mathcal{D}_M\phi\right|^2
\end{equation}
where $\gamma^M=(\gamma^\mu,\gamma^5)$, with
$\gamma^5=-i\,diag(1,-1)$, and the covariant derivative
$\mathcal{D}_M=D_M+i\vec\sigma\cdot\vec A_M$ contains the corresponding
$SU(2)_R$ gauge field. Since $A_5^2$ is an even field, it contains a
zero mode and can be given a constant background, $\langle
A_5^2\rangle$. Then the terms in the Lagrangian
\begin{equation}\label{interm}
\frac{i}{2}\bar\lambda\, i\gamma^5\,\sigma^2\,A_5^2\,\lambda+
\left| \sigma^2 A_5^2\,\phi\right|^2 
\end{equation}
generate a constant shift to the mass of all Kaluza-Klein modes as
\begin{equation}\label{shift}
\Delta m_n=\langle
A_5^2\rangle
\end{equation}
which is equivalent to the Scherk-Schwarz mechanism corresponding to the generator
$Q=\sigma^2$ and with parameter
\begin{equation}\label{parametro}
\omega=\langle
A_5^2\rangle\, R
\end{equation}

\section{Supersymmetry breaking}\label{susy}

In this section we will study the different mechanisms of
supersymmetry breaking based on the existence of extra dimensions. In
order to simplify the analysis as much as possible we will concentrate
on a five-dimensional theory, i.~e. $d=1$, where the extra dimension
is compactified on the orbifold $S^1/\mathbb{Z}_2$. In particular we
will set up the formalism for coupling the five-dimensional
super-Yang-Mills multiplets and hypermultiplets to the
$S^1/\mathbb{Z}_2$ orbifold boundaries. On this issue we will follow
the formalism introduced by Mirabelli and Peskin~\cite{MirPes}.
 
We will first consider a five-dimensional space-time with metric
$$\eta_{MN}=diag(1,-1,-1,-1,-1),$$ $M=\mu,5$, and Dirac matrices
$\gamma^M=(\gamma^\mu,\gamma^5)$ with
\begin{equation}\label{gamma}
\gamma^\mu=\left(
\begin{array}{rr}
0& \sigma^\mu\\
\bar\sigma^\mu&0
\end{array}
\right),\quad
\gamma^5=\left(
\begin{array}{rr}
-i& 0\\
0&i
\end{array}
\right)
\end{equation}
where $\sigma^\mu=(1,\vec\sigma)$ and
$\bar\sigma^\mu=(1,-\vec\sigma)$. In five-dimensional supersymmetry it
is convenient to work with Symplectic-Majorana spinors $\lambda^i$
($i=1,2$) that transform as doublets under $SU(2)_R$, see
Eq.~(\ref{SympMaj}). In particular given two Symplectic-Majorana
spinors $\lambda$ and $\chi$ they must satisfy the identity
\begin{equation}\label{identidad}
\bar\lambda^i \gamma^M\cdots\gamma^P
\chi^j=-\epsilon^{ik}\epsilon^{jl}\bar\chi^l\gamma^P\cdots\gamma^M\lambda^k
\end{equation}
that includes a minus sign from fermion interchange.

The five dimensional Yang-Mills on-shell multiplet
$\left(A_M,\Sigma,\lambda^i\right)^{\mathbb{A}dj}$ is extended to an
off-shell multiplet by adding an $SU(2)_R$ triplet of real-valued {\it
auxiliary} fields $X^a$ $(a=1,2,3)$, i.~e.
\begin{equation}\label{vectoroff}
\mathbb{V}_{off-shell}=\left(A_M,\Sigma,\lambda^i,\vec
X\right)^{\mathbb{A}dj}
\end{equation}
We will write the members of the multiplet as matrices in the adjoint
representation of the gauge group with generators $t^A$:
$\mathbb{V}=\mathbb{V}^A t^A$. The $N=1$ supersymmetric
transformations are defined by a supersymmetric parameter $\xi^i$: a
Symplectic-Majorana spinor. The supersymmetric transformations are
given by:
\begin{eqnarray}\label{susyvector}
\delta_\xi A^M &=& i\bar\xi_i\gamma^M\lambda^i\nonumber\\
\delta_\xi\Sigma &=& i\bar\xi_i\lambda^i\nonumber\\
\delta_\xi\lambda^i &=& \left(\gamma^{MN}F_{MN}-\gamma^M D_M\Sigma\right)\xi^i
-i( \vec X\cdot\vec\sigma)^{ij}\xi^j\nonumber\\
\delta_\xi X^a &=& \bar\xi_i(\sigma^a)^{ij}\gamma^MD_M\lambda^j-i
\left[\Sigma,\bar\xi_i (\sigma^a)^{ij}\lambda^j\right]
\end{eqnarray}
where $D_M\Sigma=\partial_M\Sigma-i[A_M,\Sigma]$ and
$\gamma^{MN}=[\gamma^M,\gamma^N]/4$.

The five-dimensional on-shell hypermultiplet $(A^i,\psi)$, where $A^i$
is an $SU(2)_R$ doublet and
\begin{equation}\label{Dirac}
\psi=\left(
\begin{array}{c}
\psi_L\\
\psi_R
\end{array}
\right)
\end{equation}
a Dirac spinor, is extended to the off-shell multiplet
\begin{equation}\label{hyperoff}
\mathbb{H}_{off-shell}=(A^i,\psi,F^i)
\end{equation}
where $F^i$ is an $SU(2)_R$ doublet of complex {\it auxiliary}
fields. The components of the hypermultiplet transform under $N=1$
supersymmetry as,
\begin{eqnarray}\label{susyhyper}
\delta_\xi A^i &=& -\sqrt{2}\epsilon^{ij}\bar\xi_j\psi \nonumber\\
\delta_\xi \psi &=& i\sqrt{2}\gamma^M D_M A^i\epsilon^{ij}\xi^j+
\sqrt{2}F^i\xi^i\nonumber\\
\delta_\xi F^i &=& -i\sqrt{2}\bar\xi_i\gamma^M D_M \psi
\end{eqnarray}

The previous formalism, and supersymmetric transformations
(\ref{susyvector}) and (\ref{susyhyper}), hold for a flat
five-dimensional space and also in the case of toroidal
compactifications, i.~e. in this case compactification on the circle
$S^1$. For orbifold compactifications there appear four-dimensional
fixed point branes where supersymmetry is reduced by a half. We will
discuss in this section the four-dimensional supersymmetry that
appears on the branes.

\subsection{Supersymmetry breaking by orbifolding}\label{susyorb}
In the orbifold $S^1/\mathbb{Z}_2$ there are four-dimensional branes
at the fixed points $y=0,\pi R$ where supersymmetry is reduced from
$N=2$~\footnote{$N=1$ supersymmetry in five dimensions is like $N=2$
supersymmetry in four-dimensions.} to $N=1$. In the rest of this
section we will analyze this $N=1$ supersymmetry at the
four-dimensional fixed points of the orbifold.

To project the bulk structure into the orbifold $S^1/\mathbb{Z}_2$ we
must impose the boundary conditions on fields $\phi$ as
\begin{equation}\label{bouncond}
\phi(x,-y)=Z_\phi\, \phi(x,y)
\end{equation}
where $Z_\phi=\pm 1$ are the field intrinsic parities. The parities
$Z_\phi$ must be assigned such that they leave the bulk Lagrangian
invariant. Fields $\phi$ with $Z_\phi=-1$ vanish at the walls but have
non-vanishing derivatives, $\partial_5\phi$, that can couple to them.
We will separately consider the cases of vector and hypermultiplets.

\subsubsection{Vector multiplets}
We will consider here a vector multiplet $(A_M,\Sigma,\lambda^i,X^a)$
and orbifold conditions that do not break the gauge
structure~\footnote{For a more general analysis where the gauge
structure is broken by the orbifold projection, see
section~\ref{gauge}.}. The parity assignments are chosen to be those in
table~\ref{tab1},
\begin{table}[htb]
\caption{Parities of the vector multiplet\vspace*{1pt}}
{\footnotesize
\begin{center}
\begin{tabular}{||c|c|c||}
\hline\hline
{} &{} &{} \\[-1.5ex]
{} &$Z=+1$ & $Z=-1$ \\[1ex]
\hline
{} &{} &{} \\[-1.5ex]
$A^M$ & $A^\mu$ & $A^5$\\[1ex]
$\Sigma$& &$\Sigma$\\[1ex]
$\lambda^i$ & $\lambda^1_L$ & $\lambda^2_L$\\[1ex]
$X^a$& $X^3$ & $X^{1,2}$\\[1ex]
\hline
{} &{} &{} \\[-1.5ex]
$\xi^i$& $\xi^1_L$ & $\xi^2_L$\\[1ex]
\hline\hline
\end{tabular}\label{tab1}\end{center} }
\vspace*{-13pt}
\end{table}
\noindent
where we have also included the parities of the supersymmetric
parameters. Notice that $\Sigma$ is odd and so it does not couple to the
wall, while $D_5 \Sigma=\partial_5 \Sigma$ is then even and
gauge-covariant on the wall.

From table~\ref{tab1} we can see that $\xi^1_L$ is the parameter of
the $N=1$ supersymmetry on the wall. Then supersymmetric
transformations (\ref{susyvector}) reduce on the wall $y=0$ to the
following transformations generated by $\xi^1_L$ on the even-parity
states:
\begin{eqnarray}\label{vecwall}
\delta_\xi A^\mu & = &
i\xi^{1\dagger}_L\bar\sigma^\mu\lambda^1_L-i\lambda^{1\dagger}_L\bar
\sigma^\mu\xi^1_L
\nonumber\\ \delta_\xi \lambda^1_L & = &
\gamma^{\mu\nu}F_{\mu\nu}\xi^1_L-i(X^3-\partial_5\Sigma)\xi^1_L\nonumber\\
\delta_\xi X^3 & = & \xi^{1\dagger}_L\bar\sigma^\mu D_\mu\lambda^1_L-i
\xi^{1\dagger}_L D_5\bar\lambda^2_L+h.c.\nonumber\\ \delta_\xi
\partial_5\Sigma & = & -i \xi^{1\dagger}_L D_5\bar\lambda^2_L+h.c.
\end{eqnarray}
Gathering the last two equations in (\ref{vecwall}) yields,
\begin{equation}\label{auxvec}
\delta_\xi (X^3-\partial_5\Sigma)=\xi^{1\dagger}_L\bar\sigma^\mu
D_\mu\lambda^1_L
\end{equation}
which shows that the $N=1$ vector multiplet on the brane in the
Wess-Zumino (WZ) gauge is given by $(A^\mu,\lambda^1_L,D)$ where the
auxiliary $D$-field is $D=X^3-\partial_5\Sigma$~\cite{otros}.

In this way the five dimensional action can be written as
\begin{equation}\label{Svec}
S=\int d^5x \left\{\mathcal{L}_5+\sum_i\delta(y-y_i)\mathcal{L}_{4i}\right\}
\end{equation}
where $y_i=0,\pi R$ in the present case. The bulk Lagrangian should be
the standard one for a five dimensional super-Yang-Mills theory
\begin{equation}\label{ele5vec}
\mathcal{L}_5=
{\rm tr}\,\left[-\frac{1}{2}F_{MN}^2+
(D_M\Sigma)^2+\bar\lambda i\gamma^MD_M\lambda+\vec{X}^2-\bar\lambda
[\Sigma,\lambda]\right]
\end{equation}
with ${\rm tr}\,t^At^B=\delta^{AB}/2$. The boundary Lagrangian should
have the standard form corresponding to a four-dimensional chiral
multiplet localized on the brane at $y=0$, $(\phi,\psi_L,F)$ and
coupled to the gauge $N=1$ multiplet
$(A^\mu,\lambda^1_L,X^3-\partial_5\Sigma)$.  The chiral multiplet is
supposed to transform under the irreducible representation $R$ of the
gauge group and we will call $t^A_R$ the generators of the gauge group
in the corresponding representation. The brane Lagrangian is then
written as
\begin{eqnarray}\label{ele4vec}
\mathcal{L}_4&=&{\rm tr} \left[|D_\mu\phi|^2+\bar\psi_L i
\bar\sigma^\mu D_\mu\psi_L+|F|^2\right] \nonumber\\
&-&ig\sqrt{2}(\lambda^{1A}_L\phi^\dagger
t^A_R\psi_L+\bar\psi_L t^A_R\phi\bar\lambda^1_L)+ g\,\phi^\dagger t^A_R\phi
(X^A_3-\partial_5\Sigma^A) \ .
\end{eqnarray}
The Lagrangian involving the auxiliary fields $X_3^A$ and the scalar
field $\phi$ is
\begin{equation}\label{int1}
\int d^5x \left\{\frac{1}{2} (X^{A}_3)^2+g\,\delta(y)\phi^\dagger
t^A_R\phi(X^A_3-\partial_5 \Sigma^A)\right\}
\end{equation}
Integrating out the auxiliary fields $X^A_3$ yields the boundary Lagrangian
\begin{equation}\label{boundLag}
-g\,\phi^\dagger t^A_R\phi\partial_5\Sigma^A-\frac{1}{2}g^2(\phi^\dagger
 t^A\phi)^2\delta(0)
\end{equation}
As we can see the formalism provides singular terms $\delta(0)$ on the
boundary which arise naturally from integration of auxiliary
fields. These singular terms are required by supersymmetry and they are
necessary for cancellation of divergences in the supersymmetric
limit. These terms can be formally understood as
\begin{equation}\label{delta0}
\delta(0)=\frac{1}{\pi R}\sum_{n=-\infty}^\infty 1
\end{equation}

Using Eqs.~(\ref{ele5vec}) and (\ref{boundLag}) we can write the five
dimensional Lagrangian for the $\Sigma^A$ fields as
\begin{eqnarray}\label{cuadradoV}
\mathcal{L}_5 &=&
-\frac{1}{2}(\partial_5\Sigma^A)^2-\delta(y)g\,\phi^\dagger t^A_R\phi
\partial_5\Sigma^A-\frac{1}{2}g^2 \left(\phi^\dagger t^A_R
\phi\right)^2 \delta^2(y)\nonumber\\ &=&-\frac{1}{2}\left[
\partial_5\Sigma^A+\delta(y)g\phi^\dagger t^A_R \phi\right]^2
\end{eqnarray}
We can see that the Lagrangian (\ref{cuadradoV}) is a perfect square
and the corresponding potential has a minimum at
\begin{equation}\label{sigmamin}
\Sigma^A=-\frac{1}{2}g\epsilon(y)\phi^\dagger t^A_R \phi
\end{equation}
where $\epsilon(y)$ is the sign function. We can see that if $\phi$
acquires a VEV, also $\Sigma^A$ acquires one breaking the gauge
group. The function $\Sigma^A(y)$ is an odd function and has jumps at
the orbifold fixed points. This behaviour is typical of odd functions
in orbifold backgrounds~\cite{jumps}.

\subsubsection{Hypermultiplets}

Hypermultiplets on the walls can be treated in the same way as we have
just done with vector multiplets. A hypermultiplet is defined by
$(A^i,\psi,F^i)$, where $F^i$ is a doublet of complex auxiliary
fields. A consistent set of assignments which yields $N=1$
supersymmetry on the wall is
\begin{table}[htb]
\caption{Parities of the hypermultiplet\vspace*{1pt}}
{\footnotesize
\begin{center}\begin{tabular}{||c|c|c||}
\hline\hline
{} &{} &{} \\[-1.5ex]
{} &$Z=+1$ & $Z=-1$ \\[1ex]
\hline
{} &{} &{} \\[-1.5ex]
$A^i$ & $A^1$ & $A^2$\\[1ex]
$\psi$& $\psi_L$ &$\psi_R$\\[1ex]
$F^i$ & $F^1$ & $F^2$\\[1ex]
\hline
{} &{} &{} \\[-1.5ex]
$\xi^i$& $\xi^1_L$ & $\xi^2_L$\\[1ex]
\hline\hline
\end{tabular}\label{tab2}\end{center} }
\vspace*{-13pt}
\end{table}

Similarly to vector multiplets, supersymmetry on the wall is generated
by $\xi^1_L$ and it acts on even-parity states as
\begin{eqnarray}\label{hyperwall}
\delta_\xi A^1 & = & \sqrt{2} \xi^1_L \psi_L\nonumber\\ \delta_\xi
\psi_L & = & i\sqrt{2}\sigma^\mu\partial_\mu
A^1\xi^{1*}_L-\sqrt{2}\partial_5
A^2\xi^1_L+\sqrt{2}F^1\xi^1_L\nonumber\\ \delta_\xi F^1 & = &
i\sqrt{2} \xi^{1\dagger}_L\bar\sigma^\mu\partial_\mu \psi_L+\sqrt{2}
\xi^{1\dagger}_L\partial_5\psi_R\nonumber\\
\delta_\xi\partial_5A^2& = & \sqrt{2}
\xi^{1\dagger}_L\partial_5\psi_R
\end{eqnarray}
Putting together the last two equation of (\ref{hyperwall}) leads to
\begin{equation}\label{auxhyper}
\delta_\xi(F^1-\partial_5A^2)=i\sqrt{2}
\xi^{1\dagger}_L\bar\sigma^\mu\partial_\mu \psi_L
\end{equation}
which shows that $\mathbb{A}=(A^1,\psi_L,F^1-\partial_5A^2)$
transforms as an off-shell chiral multiplet on the boundary. Notice
that, as it happened with the case of the vector multiplet, the
auxiliary field of a chiral $N=1$ multiplet on the brane does contain
the $\partial_5$ of an odd field~\cite{MirPes}.

We can now write the coupling of the bulk hypermultiplet to chiral
superfields $\Phi_0=(\phi_0,\psi_0,F_0)$ localized on the brane $y=0$
through a superpotential $W$ that depends on $\phi_0$ and the boundary
value of the scalar field $A^1$,
\begin{equation}\label{superp}
W=W(\Phi_0,\mathbb{A})
\end{equation}
The five dimensional action can then be written as in Eq.~(\ref{Svec})
with a bulk Lagrangian
\begin{equation}\label{ele5hyper}
\mathcal{L}_5=|\partial_M A^i|^2+i\bar\psi\gamma^M\partial_M\psi+|F^i|^2
\end{equation}
and a brane Lagrangian
\begin{equation}\label{ele4hyper}
\mathcal{L}_4=(F^1-\partial_5 A^2)\frac{d W}{d A^1}+h.c.
\end{equation}
Integrating out the auxiliary field $F^1$ yields
\begin{equation}\label{efe}
\bar{F}^1=-\delta(y)\frac{dW}{dA^1}
\end{equation}
and replacing it into the Lagrangian (\ref{ele5hyper}) and
(\ref{ele4hyper}) gives an action
\begin{eqnarray}\label{accionh}
S&=&\int d^5x\left\{ |\partial_M A^i|^2+i\bar\psi\gamma^M\partial_M\psi
\phantom{\frac{1^1}{1^1}}\right.\nonumber\\
&-&\left.\delta(y)\left[\left( \partial_5 A^2\frac{dW}{d A^1}+h.c.\right)+
\delta(y)\left|\frac{dW}{d A^1} \right|^2\right]
\right\}
\end{eqnarray}
where we again find a singular coupling $\delta(0)$ as required by
supersymmetry. Collecting in (\ref{accionh}) the terms where $A^2$
appears we get a potential
\begin{equation}\label{potencialA2}
V=\left| \partial_5 A^2+\delta(y)\frac{dW}{dA^1}\right|^2
\end{equation}
that is a perfect square and is then minimized for
\begin{equation}\label{minimohyper}
A^2=-\frac{1}{2}\epsilon(y) \frac{dW}{dA^1}
\end{equation}
Then if supersymmetry is spontaneously broken in the brane, i.~e. if
$$\left<\frac{dW}{dA^1}\right>\neq 0$$ then $A^2$ acquires a VEV. This
behaviour is reminiscent of a similar one in the Horava-Witten
theory~\cite{HorWit} in the presence of a gaugino condensation.

\subsection{Supersymmetry breaking by Scherk-Schwarz compactification}
\label{susySS}

In this section we will keep on considering the previous
$S^1/\mathbb{Z}_2$ orbifold and, in particular, a five dimensional
generalization of the Supersymmetric Standard
Model~\cite{PomQui,Savas}.  The gauge and Higgs sectors of the theory
will be considered to live in the bulk while chiral matter (quarks and
leptons) will be supposed to be localized on the four-dimensional
boundaries (fixed points) of the orbifold. As we have seen in the
previous section the orbifold boundary conditions break the $N=2$
supersymmetry in the bulk to $N=1$ for the zero modes and on the
branes. We will further break the residual $N=1$ supersymmetry by
using the Scherk-Schwarz boundary conditions and the global $SU(2)_R$ symmetry.

\subsubsection{Bulk breaking}

The on-shell gauge multiplet
$\mathbb{V}=(A_M,\Sigma,\lambda^i)^{\mathbb{A}dj}$ belongs to the
adjoint representation of the gauge group $SU(3)\otimes SU(2)_L\otimes
U(1)_Y$ and $i=1,2$ is an $SU(2)_R$ index. The Higgs boson fields
belong to the hypermultiplets $\mathbb{H}^a=(H^a_i,\psi^a)$, where
$a=1,2$ transforms as a doublet of the global group $SU(2)_H$. The
five-dimensional action can be written as in Eqs.~(\ref{ele5vec}) and
(\ref{ele5hyper})
\begin{eqnarray}\label{l5bulk}
\mathcal{L}_5 &=& \frac{1}{g^2}{\rm tr}\left\{-\frac{1}{2}F_{MN}^2+
(D_M\Sigma)^2+i\bar\lambda_i \gamma^MD_M\lambda^i-\bar\lambda_i
[\Sigma,\lambda^i]\right.\nonumber\\ &+&|D_M
H^a_i|^2+\bar\psi_a(i\gamma^MD_M-\Sigma)\psi^a-
(i\sqrt{2}H^{a\dagger}_i\bar\lambda_i\psi^a+h.c.)\nonumber\\
&-&\left. H^{a\dagger}_i\Sigma^2
H^a_i-\frac{g^2}{2}\left(H^{a\dagger}_i\vec{\sigma}^j_i T^A
H^a_j\right)^2\right\}
\end{eqnarray}

We now define the $\mathbb{Z}_2$ parity according to the symmetries of
the bulk Lagrangian as in table~\ref{tab3}.
\begin{table}[htb]
\caption{Parities of bulk fields\vspace*{1pt}}
{\footnotesize
\begin{center}\begin{tabular}{||ccc|ccc||}
\hline\hline
{} &{} &{} &{} &{} &\\[-1.5ex]
\multicolumn{3}{c}{$Z=+1$} & \multicolumn{3}{c}{$Z=-1$} \\[-1.5ex]
{} &{} &{} &{} &{} & \\[-1.5ex]
\hline
{} &{} &{} &{} &{} & \\[-1.5ex]
$V_\mu$ & $H^2_2$ & $H^1_1$ & $V_5,\Sigma$ & $H^2_1$ & $H^1_2$\\[1ex]
$\lambda^1_L$ & $\psi^2_L$ & $\psi^1_R$ & $\lambda^2_L$ 
& $\psi^2_R$ & $\psi^1_L$\\[1ex]
\hline\hline
\end{tabular}\label{tab3}\end{center} }
\vspace*{-13pt}
\end{table}
Columns in table~\ref{tab3} correspond to $N=1$ $D=4$ supersymmetric
multiplets. 

The Fourier expansion for $Z=\pm 1$ fields $\tilde\phi_{\pm}$ is
given by
\begin{eqnarray}\label{fourier}
\tilde\phi_+ &=&
\phi^{(0)}+\sqrt{2}\sum_{n=1}^\infty\cos\frac{ny}{R}\phi_+^{(n)}\nonumber\\
\tilde\phi_- &=&\sqrt{2}\sum_{n=1}^\infty\sin\frac{ny}{R}\phi_-^{(n)}
\end{eqnarray}
We can see from the expansion (\ref{fourier}) that the $\mathbb{Z}_2$
symmetry projects away half of the tower of KK-modes. In particular
the zero modes are the chiral $N=1$ superfields constituting the
columns of table~\ref{tab4}
\begin{table}[htb]
\caption{Zero modes\vspace*{1pt}}
{\footnotesize
\begin{center}\begin{tabular}{||c|cc||}\hline\hline
{} &{} & \\[-1.5ex]
Vector &\multicolumn{2}{c}{Chiral}\\[-1.5ex]
{} &{} & \\[-1.5ex]
\hline
{} &{} & \\[-1.5ex]
$V_\mu^{(0)}$& $H^{2\,(0)}_2$ &  $H^{1\,(0)}_1$\\[1.ex]
$\lambda^{1\,(0)}_L$ & $\psi^{2\,(0)}_L$ & $\psi^{1\,(0)}_R$\\[1.ex]
\hline\hline
\end{tabular}\label{tab4}\end{center} }
\end{table}
\noindent
while the non-zero modes are arranged in $N=2$ multiplets, as shown in
table~\ref{tab5}.
\begin{table}[htb]
\caption{Non-zero modes\vspace*{1pt}}
{\footnotesize
\begin{center}\begin{tabular}{||cc|cccc||}\hline\hline
&&&&&\\[-1.5ex]
\multicolumn{2}{c}{Vector}&\multicolumn{4}{c}{Hypermultiplets}\\[-1.5ex]
&&&&&\\ [-1.5ex]\hline 
&&&&&\\[-1.5ex]
$V_\mu^{(n)}$ & $\Sigma^{(n)}$ & $H^{1\,(n)}_1$
& $H^{1\,(n)}_2$ & $H^{2\,(n)}_1$ & $H^{2\,(n)}_2$\\[1ex]
$\lambda^{1\,(n)}_L$ &$\lambda^{2\,(n)}_L$ &
\multicolumn{2}{c}{$\psi^{1\,(n)}$}&\multicolumn{2}{c||}
{ $\psi^{2\,(n)}$}\\[1ex]
\hline\hline
\end{tabular}\label{tab5}\end{center} }
\end{table}

We will break supersymmetry by using $SU(2)_R$ as a global symmetry of
the theory, as well as $SU(2)_H$. Our definition of parity is then
\begin{equation}\label{par}
Z=\pm \sigma_3\otimes i\gamma^5
\end{equation}
where $\sigma_3$ is a matrix acting on both $SU(2)_R$ and $SU(2)_H$
indices while $\gamma^5$ is acting only on spinor indices. Similarly
the twist is defined as
\begin{equation}\label{giro}
T=e^{2\pi\,i\omega\sigma^2}
\end{equation}
as we saw in the previous section, where $\sigma_2$ is also acting on
both $SU(2)_R$ and $SU(2)_H$ indices. Notice that $Z$ and $T$ satisfy
the general condition (\ref{consistencia}). In particular we will
introduce the $\omega$-twist~\footnote{In fact one could introduce
different twists $q_R$ and $q_H$ for $SU(2)_R$ and $SU(2)_H$,
respectively, as in Ref.~\cite{PomQui,Savas}. Here we are taking for
simplicity $q_R=q_H=\omega$.} for gauginos, Higgsinos and Higgses as
in Eq.~(\ref{twistsol}):
\begin{eqnarray}\label{SStwists}
\left(
\begin{array}{c}
\lambda^1\\
\lambda^2
\end{array}
\right) &=& U(y)
\left(
\begin{array}{c}
\tilde\lambda^1\\
\tilde\lambda^2
\end{array}
\right),\quad
\left(
\begin{array}{c}
\psi^1\\
\psi^2
\end{array}
\right) = U(y)
\left(
\begin{array}{c}
\tilde\psi^1\\
\tilde\psi^2
\end{array}
\right)\nonumber\\
&& \\
\left(
\begin{array}{cc}
H^1_1 & H^1_2 \\
H^2_1 & H^2_2
\end{array}
\right) &=& U(y)
\left(
\begin{array}{cc}
\tilde{H}^1_1 & \tilde{H}^1_2 \\
\tilde{H}^2_1 & \tilde{H}^2_2
\end{array}
\right) U^{-1}(y),\quad  U(y)=e^{i\omega \sigma^2y/R}\nonumber
\end{eqnarray}
where tilded fields are periodic fields expandable in Fourier series as in
Eq.~(\ref{fourier}).

After replacing (\ref{SStwists}) in the Lagrangian (\ref{l5bulk}) we
obtain the following mass spectrum for $n\neq 0$ modes~\cite{PomQui},
\begin{eqnarray}\label{masasn}
\mathcal{L}^{(n)} &=&\frac{1}{R}\left\{
\left(
\begin{array}{cc}
\lambda^{1\,(n)} &\lambda^{2\,(n)}
\end{array}
\right)
\left(
\begin{array}{cc}
\omega & -n\\
-n & \omega
\end{array}
\right)
\left(
\begin{array}{c}
\lambda^{1\,(n)}\\
\lambda^{2\,(n)}
\end{array}
\right)
\right.
\nonumber\\
&&\nonumber\\
&+&\left.
\left(
\begin{array}{cc}
\bar\psi^{1\,(n)}_L &\bar\psi^{2\,(n)}_L
\end{array}
\right)
\left(
\begin{array}{cc}
n & -\omega\\
\omega & -n
\end{array}
\right)
\left(
\begin{array}{c}
\psi^{1\,(n)}_R\\
\psi^{2\,(n)}_R
\end{array}
\right)+h.c.
\right\}\\
&&\nonumber\\
&{\displaystyle -\frac{1}{R^2}}&
\left(
\begin{array}{cccc}
H^{(n)}_0 & H^{(n)}_2 &H^{(n)}_1 &H^{(n)}_3
\end{array}
\right)^{*}
\left(
\begin{array}{cccc}
n^2 & 0  & 0  & 0\\
  0  & n^2 &0 &0 \\
0 &0 & n^2+4\omega^2 & -4 n\omega\\
0&0 & -4 n\omega & n^2+4\omega^2\nonumber
\end{array}
\right)
\left(
\begin{array}{c}
H^{(n)}_0 \\
H^{(n)}_2 \\
H^{(n)}_1\\ 
H^{(n)}_3
\end{array}
\right)
\end{eqnarray}
where we have defined the fields $H_\mu$ as $H^a_i\equiv H_\mu
(\sigma^\mu)^a_i$ and, to simplify the notation, we skipped the tildes
on the fields. Therefore, the non-zero mode mass eigenstates
corresponding to the $n$-th level are two Majorana fermions
$[\lambda^{1\,(n)}_L\pm \lambda^{2\,(n)}_L]/\sqrt 2$ with masses
$|n\pm \omega|$, two Dirac fermions $[\psi^{1\,(n)}_L\pm
\psi^{2\,(n)}_L]/\sqrt 2$ with masses $|n\pm \omega|$, and four
complex scalars $H^{(n)}_0,\ H^{(n)}_2$ and $[H^{(n)}_1\pm
H^{(n)}_3]/\sqrt 2$ with masses $n$ and $|n\pm 2\omega|$,
respectively. Of course the mass spectrum of the fields $V_\mu^{(n)},
V_5^{(n)}, \Sigma^{(n)}$ is not modified by the Scherk-Schwarz twist. Notice that
only the Scherk-Schwarz mechanism with respect to the $SU(2)_R$ symmetry breaks
supersymmetry while the twist with respect to the $SU(2)_H$ symmetry
does just provide a supersymmetric mass.

The mass Lagrangian for zero modes
$(\lambda^{1\,(0)}_L,\psi^{2\,(0)}_L,\psi^{1\,(0)}_R,H^{(0)}_0,H^{(0)}_3)$
is,
\begin{equation}\label{masas0}
\mathcal{L}^{(0)}=\frac{\omega}{R}\left[\lambda^{1\,(0)}_L\lambda^{1\,(0)}_L+
\bar\psi^{2\,(0)}_L\psi^{1\,(0)}_R+h.c.\right]-\frac{4\omega^2}{R^2}
|H^{(0)}_3|^2
\end{equation}
and the complex scalar $H^{(0)}_0$ is massless (the Standard
Model-like Higgs).

From (\ref{masas0}) we can see how the Scherk-Schwarz mechanism breaks
supersymmetry in the zero mode sector of the five dimensional
fields. In particular it provides a mass to gauginos; in the language
of the Minimal Supersymmetric Standard Model (MSSM) we should write
\begin{equation}\label{soft12}
M_{1/2}=\frac{\omega}{R}
\end{equation}
On the other hand it gives a supersymmetric mass to Higgsinos thus
providing an extra-dimensional solution to the MSSM
$\mu$-problem. Using again the MSSM language we could also write
\begin{equation}\label{muterm}
\mu=\frac{\omega}{R}
\end{equation}

\subsubsection{Brane breaking}
As we said above we are assuming left- and right-handed quark and
lepton superfields localized on the boundary at $y=0$~\footnote{Of
course situations where only part of matter fields are localized on
the boundaries, and the rest propagating in the bulk, are easily
considered following similar lines to those found in this section and
the previous one. See section~\ref{topbreak} and
Ref.~\cite{topDel}}. The gauge superfield
$(V_\mu,\lambda^1_L,X_3-\partial_5\Sigma)$ coupling to the left-handed
quark superfield $(\tilde Q,q_L)$ gives, after eliminating the
auxiliary fields, the brane Lagrangian [see (\ref{Svec})],
\begin{eqnarray}\label{l4gauge}
\mathcal{L}_4 &=& |D_\mu\tilde Q|^2+i\bar q_L\sigma^\mu D_\mu q_L
-i\sqrt{2}(\tilde{Q}^\dagger\lambda^1_L
q_L+h.c.)-\tilde{Q}^\dagger\partial_5\Sigma \tilde{Q}\nonumber\\
&&-g^2\left[ \frac{1}{2}\left(\tilde{Q}^\dagger t^A \tilde{Q}\right)^2
  +(\tilde{Q}^\dagger t^A \tilde{Q})(H^{i\,\dagger}_a t^A
  (\sigma^3)^j_i H^a_j)\right]
\end{eqnarray}
that can be easily obtained from the Lagrangian in
Eq.~(\ref{ele4vec}). In the same way the interaction of all chiral
fermions to the bulk gauge multiplet can be computed and integration
$\int dy$ and mode decomposition yields a four dimensional Lagrangian
for Kaluza-Klein modes.

Similarly, the Yukawa couplings of the Higgs boson hypermultiplet
$(H^2_2,\psi^2_L,F^2-\partial_5 H^2_1)$ to the quark superfields
$(\tilde Q,q_L)$ and $(\tilde U,u_R)$ on the boundary can be easily
computed using (\ref{ele4hyper}). It yields,
\begin{eqnarray}\label{l4Yuk}
\mathcal{L}_4 &=& h_t \left[ H^2_2 q_L u_R+\psi^2_L(\tilde{Q}u_R+q_L \tilde{U})
-(\partial_5 H^2_1)\tilde Q \tilde U +h.c.\right]\nonumber\\
&-&\left|h_t H^2_2 \tilde Q\right|^2-\left|h_t H^2_2 \tilde U\right|^2
-\left|h_t \tilde U \tilde Q\right|^2\delta(y)
\end{eqnarray}
where $h_t$ is the top-quark Yukawa coupling. Similarly, integration
$\int dy$ and mode decomposition yields a four dimensional Lagrangian
for Kaluza-Klein modes. A similar procedure could be followed for
$N=1$ localized superfields in the leptonic sector: $(\tilde
L,\ell_L)$ and $(\tilde E,e_R)$.

\begin{figure}[ht]
\centerline{\epsfxsize=3.2in\epsfbox{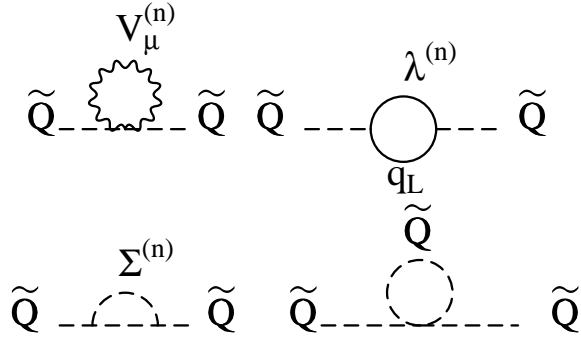}}   
\caption{Diagrams contributing to the $\tilde Q$ mass from the gauge
sector\label{figura1}}
\end{figure}

The scalar fields on the boundary (squarks and sleptons) are massless
at the tree level. Supersymmetry is broken in the bulk by gaugino
masses, Eq.~(\ref{soft12}), and transmitted to the fields on the
boundary by radiative corrections~\cite{Savas}. In particular the diagrams
for gauge interactions contributing to squark masses are exhibited in 
Fig.~\ref{figura1},
\noindent
where the Kaluza-Klein gaugino mass eigenstates are indicated by
$\lambda^{(\pm n)}=(\lambda^{1\,(n)}\pm\lambda^{2\,(n)})/\sqrt 2$.
The corresponding contribution to the squark masses is computed to be,
\begin{equation}\label{mQgauge}
m^2_{\tilde Q}=\frac{g^2 C_2(Q)}{4\pi^4}\left[\Delta m^2(0)-\Delta m^2(\omega)
\right]
\end{equation}
where $C_2(Q)$ is the quadratic Casimir of the $Q$ representation
under the gauge group~\footnote{\label{foot}We use the convention for the
generators ${\rm tr}\{t^A_R\ t^B_R\}=T(R)\,\delta^{AB}$ and $\sum_A{\rm
tr}\{t^A_R\ t^A_R\}=C_2(R)\ \Id$, where $R$ is a representation of the gauge
group. In particular if $N$ is the fundamental representation of
$SU(N)$, $T(N)=1/2$ and $C_2(N)=(N^2-1)/2N$, while for the adjoint
representation $T(\mathbb{A}dj)=C_2(\mathbb{A}dj)=N$.} and
\begin{equation}\label{defD}
\Delta m^2(\omega)=\frac{1}{2R^2}\left[Li_3\left(e^{2\pi i \omega}\right)+h.c.
\right] 
\end{equation}
where the polylogarithm functions are defined as
\begin{equation}\label{litios}
Li_n(x)=\sum_{k=1}^\infty \frac{x^k}{k^n} .
\end{equation}

\begin{figure}[ht]
\centerline{\epsfxsize=3.2in\epsfbox{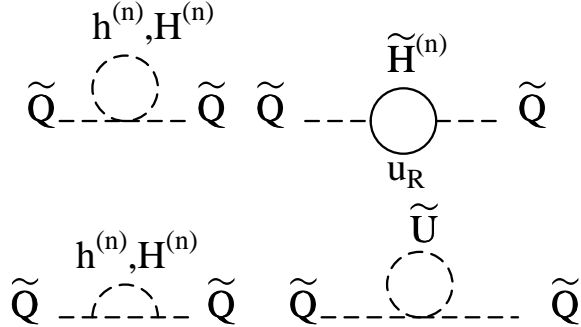}}   
\caption{Diagrams contributing to the $\tilde Q$ mass from the Higgs
sector (Yukawa couplings)\label{figura2}}
\end{figure}

The diagrams from Yukawa interactions contributing to squark masses
are given in Fig.~\ref{figura2},
\noindent
where the Kaluza-Klein mass eigenstates are defined as
$\tilde{H}^{(\pm n)}=[\psi^{1\,(n)}\pm \psi^{2\,(n)}]/\sqrt 2$, 
$h^{(|n|)}=H_0^{(n)}$, $h^{(-|n|)}=H_2^{(n)}$, $H^{(\pm n)}=[H_1^{(n)}\pm
H_3^{(n)}]/\sqrt 2$. The result is given by
\begin{equation}\label{mQYuk}
m^2_{\tilde Q}=\frac{h_t^2}{16\pi^4}\left[\Delta m^2(2\omega)+\Delta
m^2(0)-2\Delta m^2(\omega)\right]
\end{equation}

Notice that the radiative contributions to squark and slepton masses
in Eqs.~(\ref{mQgauge}) and (\ref{mQYuk}) are in all cases finite, a
feature shared by thermal masses in field theories at finite
temperature, known as Debye masses for the case of longitudinal gauge
bosons~\cite{finiteT}. The finiteness of radiative corrections to soft
masses from the tower of Kaluza-Klein modes has been challenged in
Refs.~\cite{Nilles} where it was argued that introducing a sharp
cut-off in the number of contributing Kaluza-Klein modes would restore
the typical quadratic divergences of four-dimensional field
theories. The finiteness of these radiative corrections, that was
explicitly proven to hold at two-loop by explicit
calculations~\cite{Gero2loop}, has finally been recognized to be a
robust result when one introduces a regularization that preserves the
symmetries of the five-dimensional theory, i.~e. supersymmetry and
Lorentz invariance. Furthermore explicit calculations with different
consistent regularizations all yield the {\it same finite}
result~\cite{antiNilles} therefore proving that the quadratic
divergences obtained using a sharp cut-off were an artifact of the
non-covariant regularization.  Needless to say the finiteness of the
previous results is due to supersymmetry. In fact there are indeed
quadratic divergences in the one-loop calculation that cancel by
supersymmetry.

\begin{figure}[htb]
\centerline{\epsfxsize=2.2in\epsfbox{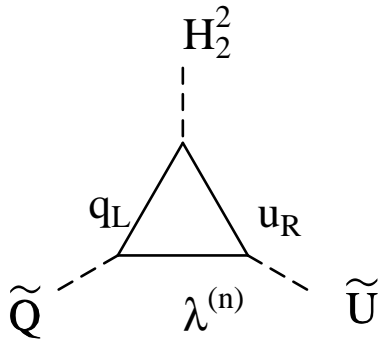}}   
\caption{Diagrams contributing to $A_t$ \label{figura3}}
\end{figure}

Finally one can compute the contribution of the Kaluza-Klein tower to
the soft breaking trilinear coupling between two boundary and one bulk
field, $A_t \tilde Q \tilde U H^2_2$. The leading contribution to the
parameter $A_t$ is provided by the exchange of gluinos as in
Fig.~\ref{figura3},
\noindent
and given by
\begin{equation}\label{At}
A_t=\frac{2\alpha_3 h_t}{3\pi^2 R}\left[i Li_2\left(e^{2\pi i \omega}\right)-i
\left(e^{-2\pi i \omega}\right)\right]
\end{equation}

Supersymmetry breaking is then gauge and Yukawa mediated to the
bosonic fields of the chiral sector localized on the branes by
radiative corrections. In this aspect the model shares common features
with any gauge mediated supersymmetry breaking model but with a very
characteristic spectrum. Electroweak symmetry breaking is also easily
triggered by radiative corrections at one-loop if the top is living in
the bulk~\cite{Savas}. In fact the tachyonic mass induced by the top
Yukawa coupling is also {\it finite}~\cite{Savas}.

The fact that the radiative contributions to soft masses are finite
does not mean that they are not sensitive to the ultraviolet cut-off
at any order of perturbation theory. It just means that {\it there is
no mass counter-term} at any order in perturbation theory. An explicit
dependence on the cut-off already appears in the two-loop correction
to the soft masses~\cite{Gero2loop}. However it comes exclusively from
the {\it wave function} renormalization and can be absorbed in the
redefinition the gauge and Yukawa couplings in the improved theory. In
fact the wave function renormalization yields a renormalization of
gauge and Yukawa couplings that contains a power-law dependence on the
cut-off. This power law renormalization~\cite{VenTay} has led to the
possibility of power-law or {\it accelerated}
unification~\cite{acelerada} in the TeV range that parallels the
ordinary logarithmic unification in the MSSM~\footnote{Scenarios where
logarithmic running leads to unification scales in the TeV range have
been discussed by Delgado and Quir\'os~\cite{acelerada} and in
Ref.~\cite{bere}.}.

\subsection{Supersymmetry breaking by Hosotani mechanism}\label{susyhosotani}

In the five dimensional formulation of local supersymmetry the
$SU(2)_R$ global supersymmetry is promoted to a local
symmetry~\cite{Zucker,Kugo,alten}. This subject is being extensively
studied at present. We will use the formulation of Ref.~\cite{Zucker}
where two multiplets are necessary to formulate off-shell five
dimensional supergravity: the minimal supergravity multiplet
$(40_B+40_F)$ and the tensor multiplet $(8_B+8_F)$. Their parities are
given in tables~\ref{tab6} and \ref{tab7}, respectively.
\begin{table}[htb]
\caption{Minimal supergravity multiplet}
{\footnotesize
\begin{center}\begin{tabular}{||c|c|c|c||}\hline\hline
{}& {} & {} & \\[-1.5ex]
Field & {}& $Z=+1$ & $Z=-1$ \\[1ex]\hline
{}& {} & {} & \\[-1.5ex]
$g_{MN}$ & graviton & $g_{\mu\nu}$, $g_{55}$ &$g_{\mu 5}$\\[1ex]
$\psi_M$ & gravitino & $\psi^1_{\mu L}$, $\psi^2_{5 L}$ &
$\psi^2_{\mu L}$, $\psi^1_{5 L}$ \\[1ex]
$B_M$ & graviphoton & $B_5$ & $B_\mu$\\[1ex]\hline
{}& {} & {} & \\[-1.5ex]
$\vec{V}_M$& $SU(2)_R$-gauge & $V_\mu^3$, $V_5^{1,2}$&$V_5^3$, $V_\mu^{1,2}$
\\[1ex]
$v^{MN}$ & antisymmetric & $v^{\mu 5}$ &$v^{\mu\nu}$ \\[1ex]
$\vec{t}$& $SU(2)_R$-triplet& $t^{1,2}$ & $t^3$\\[1ex]
$C$ & real scalar & $C$ &{}  \\[1ex]
$\zeta$ & $SU(2)_R$-doublet & $\zeta^1_L$& $\zeta^2_L$ \\[1ex]
\hline\hline
\end{tabular} \label{tab6}\end{center} }
\end{table}
\begin{table}[htb]
\caption{Tensor multiplet}
{\footnotesize
\begin{center}\begin{tabular}{||c|c|c||}\hline\hline
{}& {} &\\[-1.5ex]
Field & $Z=+1$ & $Z=-1$ \\[1ex]\hline
{}& {} &\\[-1.5ex]
$\vec{Y}$ & $Y^{1,2}$ & $Y^3$  \\[1ex]
$B_{MNP}$ & $B_{\mu\nu\rho}$ & $B_{\mu\nu 5}$\\[1ex]
$N$ & $N$ &\\[1ex]
$\rho$ & $\rho^1_L$ & $\rho^2_L$  \\[1ex]
\hline\hline
\end{tabular} \label{tab7}\end{center} }
\end{table}
Fields in the upper panel of table~\ref{tab6} are physical fields
while those in the lower panel are auxiliary fields. Fields in
table~\ref{tab7} are all of them auxiliary fields.

The $SU(2)_R$ gauge fixing is done by fixing the compensator
field~\cite{Zucker}
\begin{equation}\label{fijacion}
\vec{Y}=e^u\,\left(
\begin{array}{c}
0\\
1\\
0
\end{array}\right)
\end{equation}
that breaks $SU(2)_R\to U(1)_R=\{\sigma^2\}$. The invariant Lagrangian
$\mathcal{L}_{grav}=\mathcal{L}_{minimal}+\mathcal{L}_{tensor}$
contains the term $(1-e^u)C$ and then the equation of motion of $C$
yields $u=0$.

The auxiliary fields that are relevant for supersymmetry breaking are:
$V_5^{1,2},\ t^{1,2}$, that constitute the $F$-term of the radion
superfield~\cite{Chacko,Marti,Kaplan,GeroQui},
\begin{equation}\label{radion}
\mathcal{R}=\left[g_{55}+iB_5,\psi^2_{5 L}, V_5^1+i\,
V_5^2+4i\,(t^1+i\,t^2)\right]
\end{equation}
The relevant terms in $\mathcal{L}_{grav}$ containing these fields
are~\cite{Zucker}
\begin{eqnarray}\label{lgrav}
\mathcal{L}_{grav} &=& -\frac{i}{2}\bar\psi_P\gamma^{PMN}\mathcal{D}_M\psi_N-
\frac{1}{12}\epsilon^{MNPQR}V_M^2\partial_N B_{PQR}\nonumber\\
&+& (V^1_5)^2-12(t^1)^2-48(t^2)^2-12 N\,t^2-N^2
\end{eqnarray}
where $\gamma^{PMN}$ is the normalized antisymmetric product of gamma
matrices,
\begin{equation}\label{covderiv}
\mathcal{D}_M=D_M+i \sigma^2 V_M^2
\end{equation}
and $D_M$ is the covariant derivative with respect to local Lorentz
transformations. The field equations for the auxiliary fields yield
\begin{equation}\label{eqmov}
V^1_5=N=t^1=t^2=0
\end{equation}
while the field equation for the 3-form tensor $B_{MNP}$ gives
\begin{equation}\label{eqmovB}
\partial_{\left[M\right.}V^2_{\left.N\right]}=0 \Longrightarrow
V^2_M=\partial_M K\Longrightarrow
\left\{
\begin{array}{l}
V^2_\mu=0\ ({\rm odd\ field})\\
V_5^2={\rm constant}\ ({\rm even\ field})
\end{array}
\right.
\end{equation}
where $K$ is an odd field and the last implication is suggested by the
simplest choice
\begin{equation}\label{Kval}
K=y\,\frac{ \omega}{R}
\end{equation}
which leads to the background 
\begin{equation}\label{conectar}
V^2_5=\frac{\omega}{R}
\end{equation}
and makes the connection between the Hosotani/Wilson picture and the
Scherk-Schwarz one. In fact using the coupling of $V^2_5$ to the gravitino
field through the covariant derivative $\mathcal{D}_5$ in
(\ref{lgrav}) one obtains the gravitino mass eigenvalues for the
Kaluza-Klein modes as
\begin{equation}\label{gravmasa}
m_{3/2}^{(n)}=\frac{n+\omega}{R}
\end{equation}

An alternative choice to the odd function (\ref{Kval}) has been
proposed in Refs.~\cite{jumps,Feruglio} as
\begin{equation}\label{alternative}
K=\Lambda_1\delta(y)+\Lambda_2\delta(y-\pi R)
\end{equation}
that leads, through the covariant derivative $\mathcal D_5$ in
(\ref{lgrav}), to a localized gravitino mass term. In this case the
gravitino modes mass matrix has to be diagonalized and mass
eigenstates computed. The resulting spectrum is similar to that in
Eq.~(\ref{gravmasa}) thus proving that supersymmetry breaking by a
localized gravitino mass (that can arise e.~g. from some
non-perturbative dynamics) is equivalent to a global Scherk-Schwarz breaking, as
anticipated in Ref.~\cite{condensate}. The issue of supersymmetry
breaking by a localized mass on the brane will the subject of
section~\ref{assisted}.

Supersymmetry breaking is also manifest for gauginos and hyperscalars,
$SU(2)_R$ doublets, that interact with $\vec{V}_M$ through the
covariant derivative, see Eq.~(\ref{cinetico})
\begin{equation}\label{gaughyper}
\mathcal{L}_{matter}=\frac{i}{2}\bar\lambda\gamma^M\widehat{\mathcal{D}}_M
\lambda
+\left| \widehat{\mathcal{D}}_M A\right|
\end{equation}
where
\begin{equation}\label{covgh}
\widehat{\mathcal{D}}_M=D_M+i\,\vec\sigma\cdot\vec{V}_M
\end{equation}
After using the equations of motion $V^{1,3}_5=0$ we obtain that
$\widehat{\mathcal{D}}_M\to\mathcal{D}_M$ and the mass eigenvalues for
gauginos and hyperscalars are, as for gravitinos,
\begin{equation}\label{mattermasa}
m_{1/2}^{(n)}=m_{0}^{(n)}=\frac{n+\omega}{R}
\end{equation}
which also shows the equivalence between the Hosotani/Wilson mechanism
and the Scherk-Schwarz compactification for the matter sector. 

Supersymmetry breaking is spontaneous provided there is a (massless)
Goldstone fermion (Goldstino) ``eaten'' by the gravitino that becomes
massive in the unitary gauge. This is known as the super-Higgs
effect~\cite{superhiggs} and will be considered next.

\subsubsection{Super-Higgs effect}
\label{superHiggs}

The Goldstino is provided by the fifth component of the gravitino:
$\psi_5$. This is obvious from the local supersymmetry transformation,
\begin{equation}\label{susytrans}
\delta_\xi\psi_5=\mathcal{D}_5\xi+\cdots=i\sigma^2\,V^2_5\xi+\cdots
\end{equation}
We will now analyze the corresponding super-Higgs
effect~\cite{GeroQui}. The kinetic term for the gravitino can be
decomposed in four-dimensional and extra-dimensional components as:
\begin{eqnarray}\label{kingrav}
-\frac{i}{2}\bar\psi_M\gamma^{MNP}\mathcal{D}_N\psi_P&=&-
\frac{i}{2}\bar\psi_\mu\gamma^{\mu\nu\rho}\mathcal{D}_\nu\psi_\rho+
\frac{i}{2}\bar\psi_\mu\gamma^{\mu\nu}\gamma^5
\mathcal{D}_5\psi_\nu\nonumber\\
&-&
\frac{i}{2}\bar\psi_\mu\gamma^{\mu\nu}\gamma^5\mathcal{D}_\nu\psi_5-
\frac{i}{2}\bar\psi_5\gamma^{\mu\nu}\gamma^5\mathcal{D}_\mu\psi_\nu
\end{eqnarray}
We can now make the redefinition
\begin{equation}\label{redef}
\psi_\mu=\psi^{\prime}_\mu+\mathcal{D}_\mu(\mathcal{D}_5)^{-1}\psi_5
\end{equation}
which can be seen as a local supersymmetry transformation with
parameter $(\mathcal{D}_5)^{-1}\psi_5\equiv\xi$, gauging $\psi_5$
away. This defines a ``super-unitary'' gauge where $\psi_5$ has been
``eaten'' by the four dimensional gravitino $\psi_\mu$:
\begin{eqnarray}\label{kinunit}
-\frac{i}{2}\bar\psi_M\gamma^{MNP}\mathcal{D}_N\psi_P&=&-
\frac{i}{2}\epsilon^{\mu\nu\rho\sigma}
\bar\psi^{\prime}_\mu\gamma_\sigma\gamma^5\mathcal{D}_\nu\psi^{\prime}_\rho
\nonumber\\
&+&\frac{i}{2}\psi^{\prime}_\mu\gamma^{\mu\nu}\gamma^5
\left(\partial_5+i\,\frac{\omega}{R}\sigma^2+\cdots\right)\psi^{\prime}_\nu
\end{eqnarray}
The second term of (\ref{kinunit}) provides a mass term for the four
dimensional gravitino as
\begin{equation}\label{masagrav}
\mathcal{L}_{mass}=\frac{1}{2}\left(
\begin{array}{cc}
\psi^1_{\mu\,L}&\psi^2_{\mu\,L}
\end{array}\right)
\sigma^{\mu\nu}\left(
\begin{array}{cc}
\frac{\omega}{R} & ip_5\\
-ip_5 &\frac{\omega}{R}
\end{array}
\right)
\left(
\begin{array}{c}
\psi^1_{\nu\,L}\\
\psi^2_{\nu\,L}
\end{array}
\right)+h.c.
\end{equation}
where $p_5\equiv i\partial_5=n/R$ for a flat (unwarped) extra
dimension and $\omega$ is defined as a function of the background
field $V^2_5$ in Eq.~(\ref{conectar}). The eigenvalues of the mass
matrix (\ref{masagrav}) are $|n\pm\omega|/R$ in agreement with
Eq.~(\ref{gravmasa}).

\subsubsection{Radiative determination of the Scherk-Schwarz parameter}
\label{raddet}

The effective potential along the $V^2_5$, i.~e. $\omega$, direction
is flat and hence the Scherk-Schwarz parameter is undetermined at the
tree-level. The scale of supersymmetry breaking (i.~e. the gravitino
mass) also is undetermined at the tree-level, a situation which is
typical of no-scale models in supergravity~\cite{noscale}.

Since all mass eigenvalues, for gravitinos, gauginos and hyperscalars,
are equal to (\ref{gravmasa}) we can compute the
Coleman-Weinberg~\cite{ColeWein} one-loop effective potential in the
following way. For hyperscalars we have
\begin{equation}\label{pothyper}
V_0=\frac{2 N_H}{2}\sum_{n=-\infty}^\infty
\int\frac{d^4p}{(2\pi)^4}\log
\left[p^2+\left(\frac{n+\omega}{R}\right)^2\right]
\end{equation}
where $2 N_H=2$ (\# degrees of freedom of a complex scalar)$\times N_H$
(\# of hypermultiplets)$\times 2$ (\# scalars in one
hypermultiplet)$\times 1/2$ (orbifold reduction of \# degrees of
freedom). Similarly for gauginos we get
\begin{equation}\label{potgaug}
V_{1/2}=-\frac{2 N_V}{2}\sum_{n=-\infty}^\infty
\int\frac{d^4p}{(2\pi)^4}\log
\left[p^2+\left(\frac{n+\omega}{R}\right)^2\right]
\end{equation}
where the minus sign comes from the fermionic character of gauginos
and $N_V$ is the number of vector multiplets, while for the gravitino,
in the five dimensional harmonic gauge $\gamma^N\psi_N=0$ the
effective potential is easily worked out to be~\cite{harmonic}
\begin{equation}\label{potgrav}
V_{3/2}=-\frac{4}{2}\sum_{n=-\infty}^\infty
\int\frac{d^4p}{(2\pi)^4}\log
\left[p^2+\left(\frac{n+\omega}{R}\right)^2\right]
\end{equation}

Using techniques from field theory at finite temperature one
obtains~\cite{Savas}
\begin{equation}\label{effpotfin}
V_{eff}=\frac{3(2+N_V-N_H)}{64\pi^6R^4}\left[Li_5\left(e^{2\pi
i\omega}\right)+h.c.\right]
\end{equation}
where the polylogarithm functions are defined in Eq.~(\ref{litios}),
and $Li_5$ has a power expansion as
\begin{equation}\label{expans}
Li_5\left(e^{2\pi i\omega}\right)+h.c.=
2\zeta(5)-4\pi^2\zeta(3)\omega^2+\cdots
\end{equation}
Moreover using the expansion (\ref{expans}) it is easy to see that the global
minimum of the potential is $\omega=0$ ($\omega=1/2$) for $N_H>2+N_V$
($N_H<2+N_V$). This value can be shifted if there is another source of
supersymmetry breaking; for instance if the Scherk-Schwarz breaking is
``assisted'' by brane effects, as we will see in the next section.

\subsubsection{Brane assisted Scherk-Schwarz supersymmetry breaking}
\label{assisted}

The coupling of the minimal and tensor multiplets to the branes
permits a total Lagrangian as
\begin{equation}\label{total}
\mathcal{L}=\mathcal{L}_{grav}+2\,W\delta(y)\mathcal{L}_{brane}
\end{equation}
with~\cite{Zucker,tonis}
\begin{equation}\label{elbrana}
\mathcal{L}_{brane}=-2\,N-2\,V^1_5-12\,
t^2+\frac{1}{2}\,\bar\psi_\mu\sigma^2\gamma^{\mu\nu}\psi_\nu
\end{equation}
Where we can always assume that $W$ comes from some non-perturbative
dynamics that appears when some brane fields in the hidden sector are
integrated out. This typically happens when supersymmetry is broken by
gaugino condensation, as it is the case in
$M$-theory~\cite{condensate,Nillescond}. Of course in general one
could also introduce another term localized at the other brane, as
$2\,W^{\prime}\delta(y-\pi R)$, but we will work out here the simplest
case.

In the presence of brane terms the field equations of the auxiliary
fields $N$ and $V^1_5$ are modified to
\begin{equation}\label{auxmod}
N=-2\,W\delta(y),\quad V^1_5=2\,W\delta(y) .
\end{equation}
The presence of these VEV's modify the mass terms~\cite{GQR} for
gauginos,
\begin{equation}\label{masag}
\mathcal{M}_{1/2}=i\gamma^5\widetilde{\mathcal{D}}_5=i\gamma^5\left(
\partial_5+i\sigma^1 V_5^1+i\sigma^2 V_5^2\right)
\end{equation}
and hyperscalars, 
\begin{equation}\label{masah}
\mathcal{M}_{0}^2=\widetilde{\mathcal{D}}_5 \widetilde{\mathcal{D}}^5
\end{equation}
while the gravitino mass is directly modified by the term in
$\mathcal{L}_{brane}$: 
\begin{equation}\label{masa32}
2\,W\delta(y)\,\frac{1}{2}\, \bar\psi_\mu \gamma^{\mu\nu}\sigma^2\psi_\nu
\end{equation}

The corresponding mass eigenvalues can be seen to be modified by the
presence of $W\neq 0$ as~\cite{GQR}
\begin{equation}\label{mod}
m^{(n)}_a=\frac{n+\Delta_a(\omega,W)}{R},\quad a=0,\ 1/2,\ 3/2
\end{equation}
where the quantities $\Delta_a(\omega,W)$ have been computed to
be~\cite{jumps}
\begin{eqnarray}\label{deltas}
\Delta_{3/2}(\omega,W)&=&\arctan \sqrt{\frac{\tan^2(\omega\pi)+\tanh^2(W)}
{1+\tan^2(\omega\pi)\,\tanh^2(W)}  }\nonumber\\
&&\nonumber\\
\Delta_{1/2,0}(\omega,W)&=&\arctan \sqrt{\frac{\tan^2(\omega\pi)+\tan^2(W)}
{1+\tan^2(\omega\pi)\,\tan^2(W)}  }
\end{eqnarray}

The modified mass eigenvalues produce a modification of the effective
potential that turns out to be~\cite{GQR}
\begin{eqnarray}\label{potencialbr}
V_{\rm eff} &=& \frac{3}{36\pi^6R^4}\left[ Li_5\left(e^{2\pi i
\Delta_{3/2}(\omega,W)}\right)+h.c.\right]\nonumber\\
&&\\
&+& \frac{3(N_V-N_H)}{64\pi^6R^4}\left[ Li_5\left(e^{2\pi i
\Delta_{1/2}(\omega,W)}\right)+h.c.\right]\nonumber
\end{eqnarray}

\begin{figure}[htb]
\centerline{\epsfxsize=3.2in\epsfbox{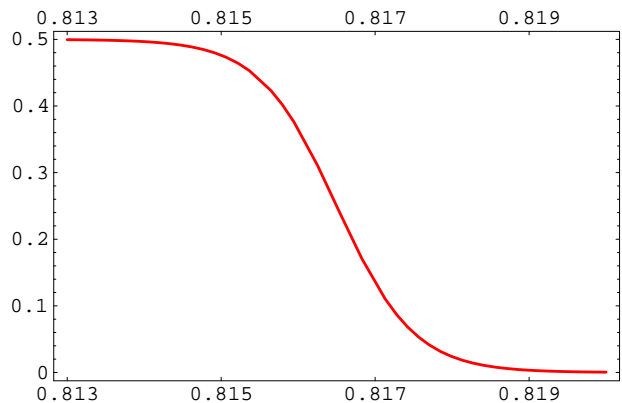}}   
\caption{Plot of the minimum of the effective potential $\omega_{min}$
as a function of $W$ for $N_V=12$ and $N_H=0$
\label{figura4}}
\end{figure}

\begin{figure}[hbt]
\centerline{\epsfxsize=3.2in\epsfbox{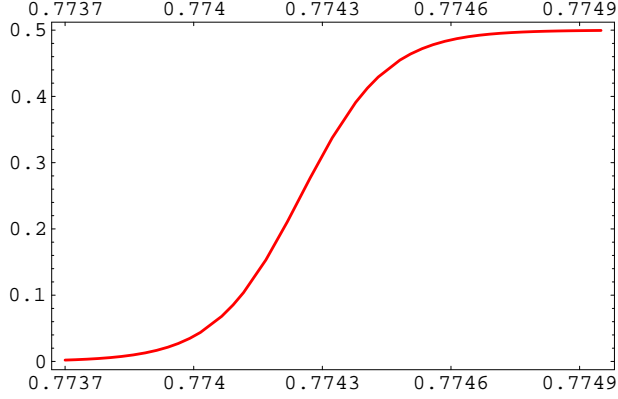}}   
\caption{Plot of the minimum of the effective potential $\omega_{min}$
as a function of $W$ for $N_V=12$ and $N_H=49$
\label{figura5}}
\end{figure}

We expect that the brane effects can modify continuously the location
of the minimum of the effective potential away from its minima for
$W=0$ at $\omega=0$ and $\omega=1/2$. We have studied numerically two
extreme cases. First of all, the case where only the gravitational and
gauge sectors are living in the bulk of the extra dimension, while
matter and Higgs fields are localized in the observable brane. This
case, that corresponds to $N_V=12$ and $N_H=0$, is shown in
Fig.~\ref{figura4}.
We can see that for $W=0$ the minimum of the potential is
$\omega_{min}=1/2$ while for $W\simgt 0.82$ it smoothly becomes
$\omega_{min}=0$ with a smooth transition region around $W\simeq
0.815$. Finally we have considered the other extreme case where all
gravitational, gauge, matter and Higgs fields are propagating in the
bulk of the extra dimensions, which amounts to considering the case
with $N_V=12$ and $N_H=49$. This case is shown in Fig.~\ref{figura5}.
We can see from Fig.~\ref{figura5} that for $W=0$ the minimum of the
potential is $\omega_{min}=0$ while for $W\simgt 0.775$ it becomes
$\omega_{min}=1/2$ with a smooth transition region around $W\simeq
0.774$.

\section{Gauge symmetry breaking}\label{gauge}

In this section we will consider the different mechanisms of gauge
symmetry breaking based on the existence of extra dimensions. As we
did in section~\ref{susy}, in order to simplify the analysis we will
concentrate on a five-dimensional theory, i.~e. $d=1$, where the extra
dimension is compactified on the orbifold $S^1/\mathbb{Z}_2$. As we
will see we can use both the orbifold (section~\ref{gaugeorb}) and Scherk-Schwarz
(section~\ref{gaugehoso}) boundary conditions to break the gauge
symmetry on the boundaries. Moreover if bulk fermions strongly coupled
to the Higgs (i.~e. the top quark) propagate in the bulk of the extra
dimensions they induce a spontaneous one-loop (Coleman-Weinberg)
finite electroweak breaking as we will describe in some detail in
section~\ref{topbreak}.

\subsection{Gauge breaking by orbifolding}\label{gaugeorb}

In section~\ref{susyorb} we have (almost) always been assuming that
the orbifold action was commuting with the gauge structure and so the
gauge group remained intact after orbifolding. This is clearly not the
most general case for the orbifold action can break the gauge group
$\mathcal{G}$ in the bulk into its subgroup $\mathcal{H}$ on the
branes. This breaking must be consistent with the orbifold action and
so it is strongly constrained. Here we will consider the case of a
general gauge group with breaking $\mathcal{G}\to\mathcal{H}$ by the
$\mathbb{Z}_2$ projection.

We consider gauge fields $A_M=A_M^A T^A$ where $T^A$ are the Lie
algebra generators of $\mathcal G$ normalized such that
\begin{equation}\label{normal}
{\rm tr}\left\{T^A T ^B\right\}=\frac{1}{2}\,\delta^{AB}
\end{equation}
with indices $M=\mu,5$ and $A=1,\cdots ,dim(\mathcal G)$. We couple to
the gauge fields Dirac spinors $\psi$ that transform in the
representation $\underline R$ [with $dim(\underline R)\equiv d_R$] of
the group $\mathcal G$.

The five dimensional action can be written as
\begin{equation}\label{accion5}
S_5=\int d^5x\ {\rm
Tr}\, \left\{-\frac{1}{2}F_{MN}F^{MN}+i\bar\psi\gamma^MD_M\psi\right\}
\end{equation}
where $F_{MN}=F_{MN}^A T^A$, $F_{MN}^A=\partial_M A_N^A-\partial_N
A_M^A+g f^{ABC} A_M^B A_N^C$ and the gauge covariant derivative is
$D_M=\partial_M-igA_M^A T_R^A$, where $T_R^A$ are Lie-algebra valued
matrices in the representation $\underline R$ of $\mathcal G$
satisfying
\begin{equation}\label{conmut}
\left[T_R^A,T_R^B \right]=if^{ABC} T_R^C .
\end{equation}

The $\mathbb Z_2$ parity assignment is defined as
\begin{eqnarray}\label{paridadV}
A_M^A(x^\mu,-y) &=& \alpha^M \Lambda^{AB}A^B_M(x^\mu,y)\nonumber\\
\psi(x^\mu,-y) &=& \lambda_R\otimes(i\gamma^5)\,\psi(x^\mu,y)
\end{eqnarray}
where $\alpha^\mu=+1$ and $\alpha^5=-1$, which just means that the
parities of $A^A_\mu$ and $A_5^A$ are opposite, and
$\lambda_R=\lambda_R^\dagger=\lambda_R^{-1}$ is a matrix acting on the
representation indices of $\psi$. Finally $\Lambda^2=1$ and so its
eigenvalues are $\pm 1$.

By imposing the requirement that under the $\mathbb{Z}_2$ action
$F_{MN}^A\to\alpha^M \Lambda^{AB} F_{MN}^B$, and that $F_{MN}^A
F_{MN}^A$ remains invariant it is straightforward to check that
$\Lambda^{AB}$ should satisfy the condition~\cite{HM-R}
\begin{equation}\label{autof}
f^{ABC}=\Lambda^{A A^\prime}\Lambda^{B B^\prime}\Lambda^{C
C^\prime}f^{A^\prime B^\prime C^\prime}
\end{equation}
which implies that the action of $\mathbb{Z}_2$ on the Lie-algebra of
$\mathcal G$ is a Lie-algebra {\it automorphism}~\footnote{A
Lie-algebra automorphism is a transformation $T^A\to \Lambda^{A
A^\prime}T^{A^\prime}$ that preserves the structure constants, i.~e. such that
$[T^{A^\prime},T^{B^\prime}]=i f^{A^\prime B^\prime
C^\prime}T^{C^\prime}$. It is easy to prove the latter equality using
Eqs.~(\ref{conmut}) and (\ref{autof}).}.

With no loss of generality we can consider the diagonal basis where
$\Lambda^{A A^\prime}=\eta^A \delta^{A A^\prime}$ with $\eta^A=\pm 1$
and the automorphism condition (\ref{autof}) takes the simpler form
\begin{equation}\label{diagonal}
f^{ABC}=\eta^A \eta^B \eta^C f^{ABC}\quad {\rm (no\ sum)}
\end{equation}

Using now the $\mathbb{Z}_2$ action (\ref{diagonal}) we can naturally
split the adjoint index $T^A$ into an unbroken part $T^a$
($\eta^a=+1$) and a broken part $T^{\hat a}$ ($\eta^{\hat a}=-1$) such
that $\mathcal{H}=\{T^a\}$,
$\mathcal{K}=\mathcal{G}/\mathcal{H}=\{T^{\hat a}\}$. In this way the
parities of gauge bosons are given in table~\ref{tab8}.
\begin{table}[htb]
\caption{Parities of gauge sector}
{\footnotesize
\begin{center}\begin{tabular}{||l|l|l||}\hline\hline
{} &{} &{} \\[-1.5ex]
EVEN & $A_\mu^a$ & $A_5^{\hat a}$\\
{} &{} &{} \\[-1.5ex]
ODD  & $A_\mu^{\hat a}$ & $A_5^a$\\[1ex]
\hline\hline
\end{tabular}\label{tab8}\end{center} }
\end{table}
Following table~\ref{tab8} only $A_\mu^a$ and $A_5^{\hat a}$ contain
zero-modes and are non-vanishing on the branes. In particular
$A_\mu^a$ are the {\it gauge bosons} on the brane corresponding to the
gauge group $\mathcal H$ and $A_5^{\hat a}$ are massless scalars on
the brane. When the latter acquire a VEV they can spontaneously break
the gauge group $\mathcal H$ to a subgroup, a mechanism known as the
Hosotani mechanism~\cite{Hosotani}. This will be discussed later on in
section~\ref{gaugehoso}.

The automorphism constraint (\ref{autof}) implies that the only
non-vanishing structure constants are $f^{abc}$ and $f^{a\hat b \hat
c}$, which strongly constrains the possible breakings. A trivial
example where we can see that the automorphism constraint is a
non-trivial one is the case $\mathcal{G}=SU(2)$. Its only
non-vanishing structure constant is $f^{123}=\varepsilon^{123}$ (and
permutations thereof). There are thus only two possibilities:
\begin{itemize}
\item
The case $a=1,2,3$, which corresponds to no breaking:
$\mathcal{H}=SU(2)$.
\item
The case $a=3$ and $\hat a=1,2$, which corresponds to the breaking
$SU(2)\to U(1)$.
\end{itemize}
In particular we can easily check that the breaking $SU(2)\to$~{\it
nothing} is {\it not allowed} by the orbifold action.

As for fermions $\psi$ in the representation $R$ of the group
$\mathcal G$, the requirement that the fermion-gauge boson coupling
$igA^A_M \bar\psi \gamma^M T^A\psi$ is invariant under the orbifold
action implies that the matrix $\lambda_R$ in (\ref{paridadV}) should
satisfy the condition~\cite{GIQ1}
\begin{equation}\label{condfer}
\lambda_R T^A_R \lambda_R=\eta^A T^A_R \Longrightarrow\left\{
\begin{array}{c}
\left[\lambda_R,T^a_R\right]=0\\
\\
\left\{\lambda_R,T^{\hat a}_R\right\}=0
\end{array}
\right.
\end{equation}

The final issue here is the gauge-fixing term and ghost Lagrangian
\begin{equation}\label{ghosts}
\mathcal L_{g.f.}+\mathcal L_{ghost}=-\frac{1}{2\xi}\partial^M
A_M^A\,\partial^M A_M^A+{\rm Tr}\,\partial^M\bar cD_M c
\end{equation}
A quick glance at the ghost-gauge field interactions shows that ghosts
$c^A$ have the same parity properties as the gauge fields $A_\mu ^A$
in (\ref{paridadV}), i.~e.
\begin{equation}\label{parghost}
c^A(x^\mu,-y)=\Lambda^{AB} c^B(x^\mu,y)
\end{equation}

Finally we would like to comment that the Lie-algebra automorphisms
come in two classes that will be subsequently analyzed:
\begin{itemize}
\item
{\it Inner automorphisms}, that can be written as a group conjugation:
$T^A\to gT^Ag^{-1}$, $g\in \mathcal G$. Inner automorphisms preserve
the rank $\Rightarrow$ $rank(\mathcal H)=rank(\mathcal G)$.
\item
{\it Outer automorphisms}, that can not be written as a group
conjugation and therefore do not preserve the rank $\Rightarrow$
$rank(\mathcal H)<rank(\mathcal G)$.
\end{itemize}

\subsubsection{Rank preserving orbifold breaking}

The case $rank(\mathcal H)=rank(\mathcal G)$ implies that none of the
generators $T_R^{\hat a}$ are diagonal and therefore $\lambda_R$ can
be chosen diagonal
\begin{equation}\label{lambdadiag}
\lambda_R=\left(
\begin{array}{cc}
\Id_{d_1} & 0\\
0& -\Id_{d_2}
\end{array}\right)
\end{equation}
where $d_1$ and $d_2$ are model dependent numbers. 

The general structure for rank preserving orbifolding is better
presented in the Cartan-Weyl basis for the generators $T^A$. In this
basis the generators are organized into the Cartan subalgebra
generators $H_i$, $i=1,\cdots, rank(\mathcal G)$, and ``raising'' and
``lowering'' generators $E_\alpha$, $\alpha=1,\cdots,\,dim(\mathcal
G)-rank(\mathcal H)$ with
\begin{equation}\label{conmC}
\left[H_i,E_\alpha\right]=\alpha_i E_\alpha
\end{equation}
where the $rank(\mathcal G)$-dimensional vector $\vec\alpha$ is the
root associated to $E_\alpha$~\cite{Slansky}. The orbifold action on
gauge fields can be written as the group conjugation
\begin{equation}\label{conj}
T^A\to g\, T^A \, g^{-1},\quad {\rm with} \quad
g=e^{-2\pi i \vec v\cdot\vec H}
\end{equation}
The orbifold action is then fully specified by the $rank(\mathcal
G)$-dimensional twist vector $\vec v$. Using now standard commutation
relations it can be verified that
\begin{eqnarray}\label{orbinner}
g\,E_\alpha\, g^{-1} &=& e^{-2\pi i \vec v\cdot\vec \alpha}E_\alpha\\
g\,H_i\, g^{-1} &=& H_i
\end{eqnarray}

\noindent
From the last equality it follows that the inner automorphism
(\ref{conj}) preserves the rank. Moreover $E_\alpha$ generators with
roots $\vec\alpha$ such that $\vec\alpha\cdot\vec v=0$ belong to the
unbroken subgroup. The problem of determining the unbroken subgroup is
thus reduced to an algebraic problem, as we will see next in some
examples.

\vspace{1.5cm}
\begin{center}
\fbox{\sc $SU(2)\to U(1)$ example}
\end{center}

\noindent
This is the simplest non-trivial example. We have already seen that
the only non-trivial $SU(2)$ breaking corresponds to $SU(2)\to U(1)$
by the $\Lambda^{AB}$ matrix
\begin{equation}\label{Lambda}
\Lambda=\left(
\begin{array}{ccc}
-1 & 0 & 0 \\
0 & -1 & 0\\
0 &0 & +1
\end{array}\right)
\end{equation}
In the Cartan-Weyl basis: $H_1=T^3$, $E_{\pm}=T^1\pm i\, T^2$, with
the non-zero roots $\alpha_\pm=\pm 1$ and the commutation rules
\begin{equation}\label{comsu2}
\left[H_1,E_\pm \right]=\pm E_\pm
\end{equation}
The unbroken $U(1)$ corresponds to the generator $H_1$. The orbifold
breaking (\ref{Lambda}) is triggered by the twist vector
$v=-\frac{1}{2}$ and the group element
\begin{equation}\label{elemg}
g=e^{i\pi T^3}
\end{equation}

Fermions in this example will be considered in the fundamental
representation $R=\underline 2$. One can easily see that
\begin{equation}\label{lambda2}
\lambda_{\underline 2}=\pm\left(
\begin{array}{cc}
1 & 0\\
0 & -1
\end{array}
\right), 
\end{equation}
i.~e. $d_1=d_2=1$ in the notation of (\ref{lambdadiag}). In fact
$\lambda_{\underline 2}$ in (\ref{lambda2}) commutes with
$T^a=T^3=\frac{1}{2}\sigma^3$ and anti-commutes with $T^{\hat
a}=\left\{ T^1=\frac{1}{2}\sigma^1, T^2=\frac{1}{2}\sigma^2\right\}$
in agreement with the general orbifold condition (\ref{condfer}). The
surviving fermions on the brane are left-handed fermions with
$U(1)$-charge equal to $+1$ and right-handed fermions with
$U(1)$-charge equal to $-1$, and so the fermion spectrum in the bulk
is non-chiral.

\vspace{.5cm}
\begin{center}
\fbox{\sc $SU(3)\to SU(2)\otimes U(1)$ example}
\end{center}

\noindent
The next to simplest example of orbifold breaking by an inner
automorphism corresponds to $SU(3)\to SU(2)\otimes U(1)$. In order to
achieve this breaking pattern we take the $\Lambda^{AB}$ matrix as
\begin{equation}\label{innersu3}
\Lambda=\left(
\begin{array}{ccc}
\Id_3 &0 &0\\
0& -\Id_4&0\\
0&0& +\Id_1
\end{array}
\right)
\end{equation}
so that the $SU(3)$ generators [see Eq.~(\ref{GM})] corresponding to
$SU(2)\otimes U(1)$ have positive parity and those corresponding to
$SU(3)/ SU(2)\otimes U(1)$ negative parity. In the Cartan-Weyl basis
we have four unbroken generators: the Cartan generators $H_1,\, H_2$,
and two other unbroken generators $E_{\pm 1}$ corresponding to the
roots $\vec\alpha_{\pm 1}=(\pm 1,0)$. The twist is $\vec v=(0,\sqrt
3)$ and so $\vec\alpha_{\pm 1}\cdot\vec v=0$. For the rest of
generators $E_{\pm 2},\, E_{\pm 3}$ we find $\exp(-2\pi i \vec
v\cdot\vec\alpha)=-1$ as expected.

Choosing fermions in the representation $R=\underline 3$ we find
$\lambda_{\underline 3}=\pm diag(+1,+1,-1)$ that commutes with the
generators of $SU(2)\otimes U(1)$ and anti-commute with those of
$SU(3)/ SU(2)\otimes U(1)$. Using now the decomposition $\underline
3=\underline 2_{\,-1} \oplus \underline 1_{\,+2}$ we obtain on the brane a
massless $SU(2)$ doublet with a given chirality and a massless $SU(2)$
singlet with opposite chirality. The spectrum is chiral and anomaly
cancellation must be enforced in this model.

A classification of breaking patterns by $\mathbb{Z}_2$ inner
automorphisms is given in table~\ref{tab9}. For more details see
Ref.~\cite{HM-R}.
\begin{table}[htb]
\caption{Breaking pattern by inner automorphisms}
{\footnotesize
\begin{center}\begin{tabular}{||c|c||}\hline\hline
{} &{} \\[-1.5ex]
$\mathcal G$ & $\mathcal H$\\[1ex]
{} &{} \\[-1.5ex]
\hline
{} &{} \\[-1.5ex]
$SU(p+q)$ & $SU(p)\otimes SU(q)\otimes U(1)$\\[1ex]
$SO(p+q)$ & $SO(p)\otimes SO(q)$, $p$ or $q$ even\\[1ex]
$SO(2n)$  & $SU(n)\otimes U(1)$\\[1ex]
$E_6$ & $SU(6)\otimes SU(2)$, $ SO(10)\otimes U(1)$\\[1ex]
\hline\hline
\end{tabular}\label{tab9}\end{center} }
\vspace*{-13pt}
\end{table}

\vspace{.5cm}
\begin{center}
\fbox{\sc Final remarks}
\end{center}

\noindent
Rank preservation by inner automorphisms holds for arbitrary
$\mathbb{Z}_N$ (abelian) orbifolds. For non-abelian orbifolds the
group element $g$ can not in general be written as $\exp\{-2\pi i \vec
v\cdot\vec H\}$ and therefore group conjugation can act non-trivially
on some $H_i$ leading to the projection of the corresponding gauge
fields and thus to rank lowering. However considering non-abelian
orbifolds is too complicated a way of achieving rank lowering for a
simpler way of doing it does exist in the $\mathbb{Z}_N$ orbifolds
when the discrete group is not realized as a group conjugation, as we
will see in the next section.

\subsubsection{Rank lowering}

If the discrete group in Eq.~(\ref{paridadV}) {\it can not} be
realized as the group conjugation (\ref{conj}) its action on the Lie
algebra is called an outer automorphism and leads to rank
reduction. In short, outer automorphisms are structure constant
preserving linear transformations of the generators that can not be
written as a group conjugation. For any given Lie algebra there is
only a limited number of outer automorphisms depending on the
symmetries of the Dynkin diagrams.

Reduced rank implies that some of the $T_R^{\hat a}$ are diagonal and
for them the identity $\{\lambda_R,T_R^{\hat a}\}=0$ can never be
satisfied if $\lambda_R$ is diagonal. The most interesting case is
\begin{equation}\label{nonconj}
\Lambda^{AB}T_R^B=-\left(T_R^A\right)^T
\end{equation}
which of course preserves the structure constants. This is an outer
automorphism that produces the breaking $SU(N)\to SO(N)$. In this case
we must find a non-diagonal (and unitary) matrix in the
$R$-representation, $\lambda_R$, that solves Eqs.~(\ref{condfer}),
i.~e.
\begin{equation}\label{condfer2}
\lambda_R T_R^A\lambda_R=-\left(T_R^A\right)^T
\end{equation}
This identity can only be satisfied if $R$ is a real
representation. In fact one can always choose $R=\mathcal R\oplus
\bar{\mathcal R}$ where $\mathcal R$ is a non-real representation with
generators
\begin{equation}\label{Tred}
T_R^A=\left(
\begin{array}{cc}
T_{\mathcal R}^A & 0 \\
0 & -\left(T_{\mathcal R}^A\right)^T
\end{array}
\right)
\end{equation}
which implies that $\lambda_R$ takes the block form
\begin{equation}\label{block}
\lambda_R=\left(
\begin{array}{cc}
0 & \Id_{\mathcal R} \\
\Id_{\mathcal R}&0
\end{array}
\right)
\end{equation}
and so it has $d_{\mathcal R}$ eigenstates $(\hat e_i,\hat e_i)$ with
positive parity and $d_{\mathcal R}$ eigenstates $(\hat e_i,-\hat
e_i)$ with negative parity, where $\hat e_i$ denotes the usual unit
vector with a unit in the $i$-th entry and zero otherwise.

The zero-mode spectrum resulting from this action will always be
vector like and therefore anomaly-free. To produce a chiral theory the
simplest possibility would be to add chiral, anomaly-free, matter of
the $\mathcal H$ subgroup localized on the brane where $\mathcal G \to
\mathcal H$ via the outer automorphism. Explicit examples will be
presented next.

\vspace{.5cm}
\begin{center}
\fbox{\sc $U(1)\to nothing\quad$ example}
\end{center}

\noindent
The simplest example is $U(1)$ in the bulk breaking down to nothing on
the branes. The choice that performs this breaking is $\Lambda=-1$,
that is a simple realization of the equation $\Lambda\,T=-T^T$, with
$T=1$. Charged fermions are necessarily accompanied by oppositely
charged partners. The fermionic zero mode spectrum is a vector-like
pair of Weyl fermions that can be assembled into a Dirac fermion.

\vspace{.5cm}
\begin{center}
\fbox{\sc $SU(3)\to SO(3)$ example}
\end{center}

\noindent
The second simplest example of an automorphism with rank breaking
action is $SU(3)\to SO(3)$. Solving equation (\ref{nonconj}) one must
give a positive parity to antisymmetric generators and a negative
parity to symmetric ones. In particular the generators for the $SU(3)$
group are given by the Gell-Man matrices:
\begin{eqnarray}\label{GM}
\lambda^1 &=& \frac{1}{2}
\left(
\begin{array}{ccc}
0 & 1 & 0\\
1 & 0 & 0 \\
0 & 0 & 0 \\
\end{array}
\right)
\quad
\lambda^2 = \frac{1}{2}
\left(
\begin{array}{ccc}
0 & -i & 0\\
i & 0 & 0 \\
0 & 0 & 0 \\
\end{array}
\right)\quad
\lambda^3 = \frac{1}{2}
\left(
\begin{array}{ccc}
1 & 0 & 0\\
0 & -1 & 0 \\
0 & 0 & 0 \\
\end{array}
\right)\quad
\nonumber\\
\lambda^4 &=& \frac{1}{2}
\left(
\begin{array}{ccc}
0 & 0 & 1\\
0 & 0 & 0 \\
1 & 0 & 0 \\
\end{array}
\right)\quad
\lambda^5 = \frac{1}{2}
\left(
\begin{array}{ccc}
0 & 0 & -i\\
0 & 0 & 0 \\
i & 0 & 0 \\
\end{array}
\right)
\\
\lambda^6 &=& \frac{1}{2}
\left(
\begin{array}{ccc}
0 & 0 & 0\\
0 & 0 & 1 \\
0 & 1 & 0 \\
\end{array}
\right)
\quad
\lambda^7 = \frac{1}{2}
\left(
\begin{array}{ccc}
0 & 0 & 0\\
0 & 0 & -i \\
0 & i & 0 \\
\end{array}
\right)
\quad
\lambda^8 = \frac{1}{2\sqrt 3}
\left(
\begin{array}{ccc}
1 & 0 & 0\\
0 & 1 & 0 \\
0 & 0 & -2 \\
\end{array}
\right)
\nonumber
\end{eqnarray}
This results in the choice 
\begin{equation}\label{choice}
\Lambda=diag(-1,+1,-1,-1,+1,-1,+1,-1)
\end{equation}
The positive parity generators $(\lambda^2,\lambda^5,\lambda^7)$ form
the fundamental representation of $SO(3)$. Matter in the $3\oplus
\bar{\underline 3}$ representation transforms in the $3\oplus 3$ of
$SO(3)$. The parity of these states is
\begin{equation}\label{par3}
\lambda_{3\oplus 3}=\left(
\begin{array}{cc}
0 & \Id_3 \\
\Id_3 & 0
\end{array}\right)
\end{equation}
which, after diagonalization yields $3$ even and $3$ odd states. The
fermionic zero mode spectrum is then an $SO(3)$ triplet of Weyl
fermions plus their vector-like partners.

A classification of breaking patterns by $\mathbb{Z}_2$ outer
automorphisms is given in table~\ref{tab10}. For more details see
Ref.~\cite{HM-R}.
\begin{table}[htb]
\caption{Breaking pattern by outer automorphisms}
{\footnotesize
\begin{center}\begin{tabular}{||c|c||}\hline\hline
{} &{} \\[-1.5ex]
$\mathcal G$ & $\mathcal H$\\[1ex]
{} &{} \\[-1.5ex]
\hline
{} &{} \\[-1.5ex]
$SU(p)$ & $SO(p)$\\[1ex]
$SO(p+q)$ & $SO(p)\otimes SO(q)$\\[1ex]
$SU(2p)$  & $Sp(p)$\\[1ex]
$E_6$ & $Sp(4)$, $ F_4$\\[1ex]
\hline\hline
\end{tabular}\label{tab10}\end{center} }
\vspace*{-13pt}
\end{table}

\subsection{Gauge breaking by the Hosotani mechanism}\label{gaugehoso}

In the previous section we have seen how a gauge symmetry $\mathcal G$
can be broken down to a subgroup $\mathcal H$ by the orbifold action
on the branes. The original gauge bosons $A_\mu^A$ split into gauge
bosons on the brane $A_\mu^a$ and coset ``gauge bosons'' $A_\mu^{\hat
a}$ that are parity-odd and have no zero modes. Similarly the
extra-dimensional components $A_5^A$ of the higher-dimensional gauge
bosons split into odd scalars $A_5^a$ and even scalars $A_5^{\hat
a}$. In particular the scalars $A_5^{\hat a}$ propagate in the brane
and have massless zero modes that belong to non-trivial
representations of $\mathcal H$. For instance in the simple case
presented in section~\ref{gaugeorb}, $\mathcal G=SU(3)$ and $\mathcal
H=SU(2)\otimes U(1)$, we have that under $SU(3)\to SU(2)\otimes U(1)$
the branching ratio of the adjoint representation $8$ is
\begin{equation}\label{ratio}
8=3_0\oplus 1_0\oplus 2_Y\oplus \bar 2_{-Y}
\end{equation}
and then the scalars $A_5^A$ along the coset direction $SU(3)/
SU(2)\otimes U(1)$ are $SU(2)$ doublets.
 
The scalars $A_5^{\hat a}$ do not have any tree-level mass since it is
forbidden by the higher-dimensional gauge invariance. In flat infinite
five-dimensional space-time the gauge invariance in the
higher-dimensional space-time would keep on protecting the appearance
of a mass for $A_5^{\hat a}$ at any order in perturbation
theory. However in {\it compact spaces}, as it is the case of the
$\mathbb{Z}_2$ orbifold, the masses of $A_5^{\hat a}$ are protected
only up to {\it finite terms} of order $1/R^2$ (that go to zero in the
``infinite'' limit $R\to \infty$). These mass corrections are similar
to the thermal masses in field theory at finite
temperature~\cite{finiteT} that appear from radiative corrections
$\sim g\, T$: the so-called Debye masses for longitudinal photons and
gluons.

At tree-level the gauge symmetry $\mathcal H$ on the brane is
spontaneously broken in the presence of the background field
$A_5^{\hat a}$. However the VEV of $A_5^{\hat a}$ is undetermined at
the classical level. When radiative corrections are considered a
tachyonic mass for some $A_5^{\hat a}$ fields can be generated,
denoting an instability along the $A_5^{\hat a}$ direction whose
minimum can be found using effective potential techniques. In this
case the VEV $\langle A_5^{\hat a}\rangle$ is fixed and some gauge
bosons in $\mathcal H$ become massive. This spontaneous breaking of
$\mathcal H$ to a subgroup by $\langle A_5^{\hat a}\rangle$ is called
Hosotani breaking~\cite{Hosotani} and it has been discussed several
times along these lectures. A sufficient condition for Hosotani
breaking is that $A_5^{\hat a}$ acquires a tachyonic
mass~\footnote{There are some exceptions to this rule when $\langle
A_5^{\hat a}\rangle=1/2$ and the gauge group $\mathcal H$ is
unbroken~\cite{japanese}.}. The requirements for this sufficient
condition to hold can be established in a general theory with a gauge
group $\mathcal G$ in the bulk and an arbitrary fermionic
content. Before showing them we will remind that such a condition (a
tachyonic mass for $A_5^{\hat a}$) although sufficient is by no means
necessary since the effective potential could have a minimum at the
origin and another deeper (global) minimum located at some value
$A_5^{\hat a}\neq 0$. In that case we must rely on the cosmological
evolution leading the field towards the global minimum, or otherwise
if the life-time of the transition from the local minimum at the
origin to the global minimum is shorter than the age of the present
universe.

\begin{figure}[htb]
\begin{center}
\begin{picture}(120,60)(0,0)
\Photon(0,0)(120,0){2.5}{15.5}
\PhotonArc(60,20)(20,-90,250){2.5}{15.5} 	
\Text(20,13)[r]{$A_5^{\hat a}$}
\Text(100,13)[l]{$A_5^{\hat b}$}
\Text(60,52)[c]{$A_{\mu,5}^{c,\hat c}$}
\end{picture}
\quad 
\begin{picture}(120,60)(0,0)
\PhotonArc(60,20)(20,0,360){2.5}{15.5}
\Photon(0,20)(40,20){2.5}{5.5}
\Photon(80,20)(120,20){2.5}{5.5}	
\Text(20,33)[r]{$A_5^{\hat a}$}
\Text(100,33)[l]{$A_5^{\hat b}$}
\Text(60,52)[c]{$A_{\mu,5}^{c,\hat c}$}
\end{picture}

\vspace{.5cm}
\begin{picture}(120,60)(0,0)
\DashCArc(60,20)(20,0,360)3
\Photon(0,20)(40,20){2.5}{5.5} 
\Photon(80,20)(120,20){2.5}{5.5}
\Text(20,13)[r]{$A_5^{\hat a}$}
\Text(100,13)[l]{$A_5^{\hat b}$}
\Text(60,51)[c]{$c^{c,\hat c}$}
\end{picture}\quad\quad
\begin{picture}(120,60)(0,0)
\ArrowArcn(60,20)(20,0,180)
\ArrowArcn(60,20)(20,180,360)
\Photon(0,20)(40,20){2.5}{5.5}
\Photon(80,20)(120,20){2.5}{5.5}
\Text(20,33)[r]{$A_5^{\hat a}$}
\Text(100,33)[l]{$A_5^{\hat b}$}
\Text(60,51)[c]{$\psi$}
\end{picture}
\end{center}
\caption{The one-loop diagrams contributing to mass and wave function 
renormalization}
\label{figura6}
\end{figure}
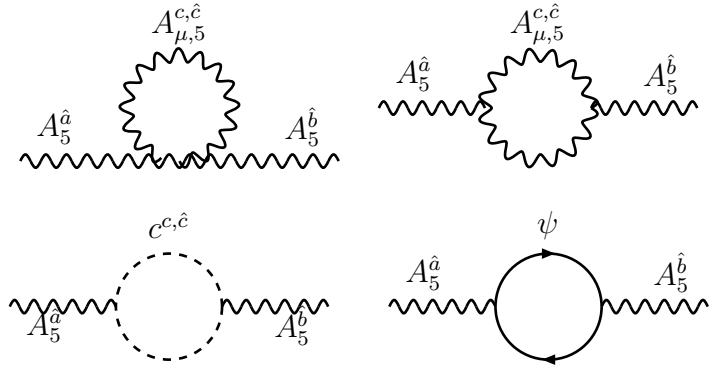

The diagrams contributing to the mass of $A_5^{\hat a}$ are shown in
Fig~\ref{figura6}.
This calculation has been done in Refs.~\cite{GIQ1,Schmaltz} leading to
the result for the $A_5^{\hat a}$ mass
\begin{equation}\label{masaH}
m^2_{\hat a}=\frac{3 g^2}{32\pi^4 R^2}\zeta(3)\left[3 C_2(\mathcal
G)-4 T(R) N_f\right]
\end{equation}
where $C_2(\mathcal G)$ is the quadratic Casimir of (the adjoint
representation of) $\mathcal G$, e.~g. $C_2[SU(N)]=N$ and $T(R)$ is the
Dynkin index of the representation $R$ satisfying
\begin{equation}\label{Dynkin}
d_R\, C_2(R)=T(R)\, d_{\mathcal G}
\end{equation}
[See footnote~\ref{foot} for more details.]

From the expression~(\ref{masaH}) we see that for models with
\begin{equation}\label{condi}
\frac{C_2(\mathcal G)}{T(R)}<\frac{4}{3}\, N_f
\end{equation}
$m_{\hat a}^2<0$ and therefore $\mathcal H$ could be radiatively
broken. To be more precise we have to look at the full effective
potential
\begin{equation}\label{potenc}
V_{eff}=\frac{1}{128\pi^6 R^4} {\rm Tr}\left[V(r_F)-V(r_B)  \right]
\end{equation}
with $r_{B,F}=\exp[2\pi i q_{B,F}(\omega)]$, $q_{B,F}$ being the shifts
in boson and fermion Kaluza-Klein masses according to
\begin{equation}\label{corrim}
m_{B,F}^{(n)}=\frac{n+q_{B,F}(\omega)}{R},\quad n\in\mathbb Z
\end{equation}
where the parameter $\omega$ measures the VEV of $A_5^{\hat a}$ as 
\begin{equation}\label{vev}
\omega=\langle A_5^{\hat a}\rangle R
\end{equation}
which provides a Scherk-Schwarz-like breaking. 

The structure of $V_{eff}$ depends on the gauge and matter
structure. A particularly simple case is when there are $N_f$ fermions
in the adjoint representation. Then $T(\mathbb Adj)=C_2(\mathcal G)$
and the symmetry is fully determined by the factor $3-4\, N_f$ in
front of the effective potential.

Notice that the radiative mass (\ref{masaH}) is finite and we want to
conclude this section with a few comments about this finiteness. The
field $A_5^{\hat a}$ is a scalar, in a given representation of the
gauge group $\mathcal H$, when it propagates on the brane. As such it
seems that its radiatively corrected mass should not be protected from
quadratic divergences (let us remind that the theory is not
supersymmetric) by the $\mathcal H$ invariance and furthermore that
radiative corrections localized on the brane should indeed be
quadratically divergent. However the calculation leading to the result
in (\ref{masaH}) contains {\it both} bulk and brane effects and
nevertheless it has been explicitly found to be finite. Is this fact
reflecting some additional symmetry or is there an accidental
cancellation of quadratic divergences? The answer to those questions
were given in Refs.~\cite{GIQ2,GIQ3} where the invariance on the
brane, the residue of the gauge invariance in the bulk, was fully
identified. In fact it was proven that there is, apart from the
$\mathcal H$ gauge invariance on the brane, an infinite set of
symmetries that do not close in a group but prevent many explicit
renormalization terms allowed by $\mathcal H$ invariance.  In
particular there is a residual shift symmetry~\cite{GIQ2} as
$A_5^{\hat a}\to A_5^{\hat a}+\partial_5 \xi^{\hat a}$, where
$\xi^{\hat a}$ is an odd gauge parameter, that prevents the appearance
in a five-dimensional gauge theory of brane mass terms for $A_5^{\hat
a}$ at any order in perturbation theory. This striking result is a
general feature of five-dimensional theories, where the non-appearance
of quadratically divergent masses on the brane for the extra
dimensional components of the five-dimensional gauge bosons is due to
the original higher-dimensional gauge invariance. However this result
can not be straightforwardly generalized to theories with more than
one extra dimension, although some finite results have been found by
explicit calculations in six-dimensional theories~\cite{NJP,Csaki}. In
fact we have proven~\cite{GIQ3} that in theories with six or more
dimensions there can exist quadratic divergences to the mass of extra
dimensional gauge bosons localized on the branes only provided that
there are $U(1)$ subgroups in $\mathcal H$ that were not already
present in $\mathcal G$. These quadratic divergences are associated to
the possible existence of tadpoles for the even fields $F_{ij}^a$ a
phenomenon similar, but not quite coincident, to that of localized
anomalies on branes~\cite{anomalies,6Dorbi}. In particular models in
six or more dimensions (unlike in five-dimensional theories) the
cancellation of brane quadratic divergences has to be enforced, as the
cancellation of global and local anomalies should be. Particular
examples with finite renormalized masses~\cite{NJP,Csaki} satisfy the
condition for cancellation of quadratic divergences, as it should
happen.

To conclude this section, we have seen that supersymmetry is not the
only symmetry that can protect the scalar masses from quadratic
radiative corrections. If gauge symmetry breaking proceeds through the
Hosotani breaking the higher-dimensional gauge invariance can do the
job thus providing a possible non-supersymmetric alternative solution
to the hierarchy problem.

\subsection{Top assisted electroweak breaking}\label{topbreak}
\begin{figure}[htb]
\begin{center}
\begin{picture}(120,60)(0,0)
\ArrowArcn(60,20)(20,0,180)
\ArrowArcn(60,20)(20,180,360)
\DashLine(0,20)(40,20){5} 
\DashLine(80,20)(120,20){5} 
\Text(20,33)[r]{$H_2$}
\Text(100,33)[l]{$H_2$}
\Text(60,52)[c]{$t_L$}
\Text(60,10)[c]{$t_R$}
\Text(180,21)[r]{+\ SUSY=0}
\end{picture}
\end{center}
\caption{The one-loop renormalization of the Higgs mass}
\label{figura7}
\end{figure}
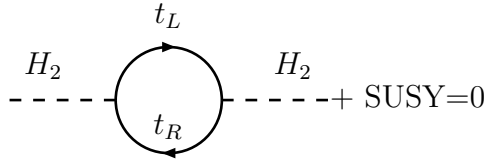

In this section we have been concentrating up to now in
non-supersymmetric models where the finiteness of the Hosotani
breaking was protected by the underlying higher-dimensional gauge
theory. Another possibility for mass protection is when the higher
(e.~g. five) dimensional theory is supersymmetric, as in
section~\ref{susy}. In that case, as we discussed there, supersymmetry
can be broken by Scherk-Schwarz compactification. In particular, in the model we
discussed in section~\ref{susySS} the gauge and Higgs hypermultiplets
were living in the bulk of the fifth dimension while quark and lepton
chiral $N=1$ supermultiplets were localized on the brane at
$y=0$. Since the brane is supersymmetric the contributions to the
Higgs mass from the top quark Yukawa coupling, see Fig.~\ref{figura7},
vanish at tree level,
while the contributions to the squared Higgs mass from gauge
interactions are positive $\Longrightarrow$ electroweak symmetry is
unbroken at one-loop.

A way of putting a remedy to this situation is by allowing the top
quark to propagate in the bulk~\cite{Savas,topDel}. Now to have
superpotential interactions on the brane as in the MSSM,
\begin{equation}\label{superpot}
W=\left[ H_t Q_L U_R H_2+h_b Q_L D_R H_1+h_\tau L_L H_1 E_R+\mu
H_1\cdot H_2\right]\,\delta(y)
\end{equation}
where we are using standard MSSM notation by which superfields are
denoted by capital Roman characters, we need to enforce orbifold
invariance on it. Since the gauge group is unbroken by the orbifold
action, $T^A=\{T^a \}$, $T^{\hat a}=\{0\}$, the conditions
(\ref{condfer}) are trivially satisfied by the identity matrices,
$\lambda_R=\Id$, and the orbifold conditions for bulk fermions are
$\psi(x^\mu,-y)=i\gamma^5 \psi(x^\mu,y)$, see Eq.~(\ref{paridadV}),
which means in Weyl components
\begin{eqnarray}\label{Weylcom}
\psi_L(x^\mu,-y)&=&-\psi_L(x^\mu,y)\nonumber\\
\psi_R(x^\mu,-y)&=&+\psi_R(x^\mu,y)
\end{eqnarray}
under which the theory is invariant. We can then consistently choose
as fields propagating in the bulk:
\begin{itemize}
\item
$\mathbb V$: the gauge sector multiplets
\item
$\mathbb U_R$, $\mathbb D_R$, $\mathbb E_R$: ``right-handed''
hypermultiplets. Their massless zero modes surviving the $\mathbb
Z_2$-projection are $(\tilde u^{(0)}_R,u^{(0)}_R)$, $(\tilde
d^{(0)}_R,d^{(0)}_R)$ and $(\tilde \ell^{(0)}_R,\ell^{(0)}_R)$.
\end{itemize}
The localized fields are the $SU(2)_L$ doublets: $H_{1,2}$,
$Q_L=(\tilde q_L,q_L)$, $L_L=(\tilde \ell_L,\ell_L)$. Their fermionic
modes will cancel the global anomalies generated by
$u_R^{(0)},\,d_R^{(0)},\,\ell_R^{(0)}$. In other words $q_l,\, \ell_L$
are the $\mathbb Z_2$-partners of
$u_R^{(0)},\,d_R^{(0)},\,\ell_R^{(0)}$ that are needed to cancel the
anomalies, which means that they should have an intrisic odd
parity~\cite{DelQuiPLB}:
\begin{equation}\label{intrisica}
\mathbb Z_2:\quad q_L\to -q_L,\quad \ell_L\to -\ell_L
\end{equation}
in the underlying theory that provides the localized states which are
necessary to achieve anomaly cancellation

Since only even $(E)$ components of bulk fields couple to the brane
[e.~g.~$U_R$] and brane localized states are $\mathbb Z_2$-odd $(O)$
[e.~g.~$H_2,\, Q_L$] the only orbifold allowed brane couplings are
$EEE$ and $EOO$. Now while $EEE$ couplings are suppressed in field
theory, since no wave function of the involved fields is concentrated
on the brane, the natural Yukawa couplings localized on the brane are
$EOO$: for instance the coupling $h_t Q_L U_R H_2$ that we are
considering in (\ref{superpot}). This justifies our choice of matter
content. Of course there are other matter contents which are also
consistent with the orbifold invariance and lead to somewhat different
phenomenological behaviours. The choice made here is just the most
convenient one for phenomenological purposes.

Given the previous matter content, the Higgs potential along the
neutral components of $H_1$ and $H_2$ is,
\begin{eqnarray}\label{potHiggs}
V(H_1,H_2) &= & m_1^2 |H_1|^2+m_2^2 |H_2|^2+m_3^2(H_1\cdot H_2+h.c.)
\nonumber\\
&+&
\frac{g^2+g^{\prime 2}}{8}(|H_1|^2-|H_2|^2)^2+\lambda_t |H_2|^2+
\lambda_b |H_1|^2
\end{eqnarray}
where the tree-level relations
\begin{equation}\label{tlrel}
M_1^2=m_2^2=\mu^2,\quad m_3^2=0,\quad \lambda_t=\lambda_b=0
\end{equation}
are satisfied. 

Radiative corrections will alter these relations since supersymmetry
is broken by the Scherk-Schwarz mechanism as we described in
section~\ref{susySS}. To compute radiative corrections we first
decompose even and odd fields into modes as in Eq.~(\ref{fourier}) and
change basis using $\phi_+^{(-n)}\equiv \phi_+^{(n)}$ and
$\phi_-^{(-n)}\equiv -\phi_-^{(n)}$ which gives the Fourier decomposition
for periodic fields as
\begin{eqnarray}\label{fouriershift}
\tilde\phi_+ &=&
\sum_{n=-\infty}^\infty\cos\frac{(n+\omega)y}{R}\phi^{(n)}\nonumber\\
\tilde\phi_-
&=&\sum_{n=-\infty}^\infty\sin\frac{(n+\omega)y}{R}\phi^{(n)}
\end{eqnarray}

In particular one-loop corrections to $m_1^2$ and $m_2^2$ can be
obtained using the results of section~\ref{susySS}.  They yield
\begin{eqnarray}\label{masas12}
m_2^2 &=& \mu^2+\frac{6 h_t^2+3g^2}{16\pi^4} \left[\Delta m^2(0)
-\Delta m^2(\omega)\right]\nonumber\\
m_1^2 &=& \mu^2+\frac{6 h_b^2+3g^2}{16\pi^4} \left[\Delta m^2(0)
-\Delta m^2(\omega)\right]
\end{eqnarray}
where $\Delta m^2(\omega)$ was defined in Eq.~(\ref{defD}) and we are
neglecting $\mathcal O(g^\prime)$-corrections. The top and bottom
Yukawa couplings are defined as
\begin{equation}\label{yukawas}
h_t=\frac{m_t}{v}\sqrt{\frac{1+\tan^\beta}{\tan^2\beta}},\quad
h_b=\frac{m_b}{v}\sqrt{1+\tan^2\beta}
\end{equation}
where $\tan\beta\equiv v_2/v_1$, $v=\sqrt{v_1^2+v_2^2}=174$
GeV. Obviously $h_b$ is only relevant for $\tan\beta\gg 1$.

\begin{figure}[htb]
\begin{center}
\begin{picture}(120,60)(0,0)
\CArc(60,20)(20,0,180)
\CArc(60,20)(20,180,360)
\DashLine(0,20)(40,20){5} 
\DashLine(80,20)(120,20){5} 
\Text(20,33)[r]{$H_1$}
\Text(100,33)[l]{$H_2$}
\Text(60,52)[c]{$\lambda^{(n)}$}
\Text(60,10)[c]{$\mu$}
\Text(40,5)[r]{$\widetilde H_1$}
\Text(80,5)[l]{$\widetilde H_2$}
\Vertex(60,0){2}
\Text(180,21)[r]{+\ SUSY=0}
\end{picture}
\end{center}
\caption{The one-loop renormalization of $m_3^2$}
\label{figura8}
\end{figure}
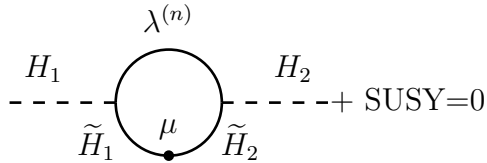

The mass term $m_3^2$ is generated by the diagram of Fig.~\ref{figura8}
and the resulting contribution is
\begin{equation}\label{renm32}
m_3^2=\frac{3 g^2\omega \mu}{512\pi^2 R}\left[i Li_2\left(e^{2\pi i
\omega}\right)+h.c.\right]\equiv \mu B
\end{equation}
From (\ref{renm32}) it is easy to check that $m_3^2$ goes to zero both
in the supersymmetric limit $(\omega\to 0)$ and when $\mu=0$, as it
should. It is also zero for the degenerate case $\omega=1/2$,
reflecting the fact that in that case the gauginos $\lambda^{(n)}$ are
Dirac fermions.

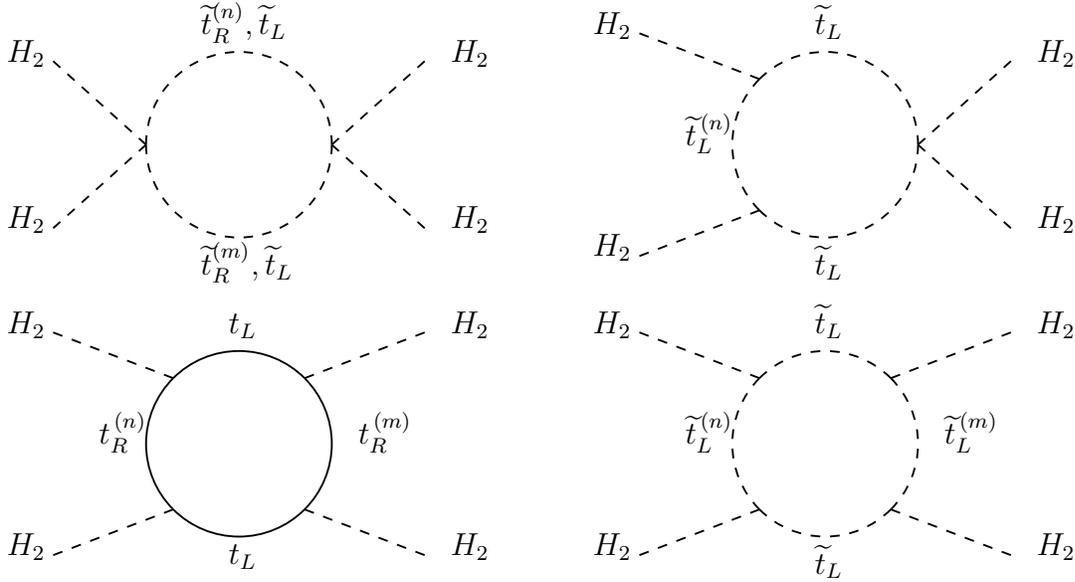
\begin{figure}[H]
\setlength{\unitlength}{.7pt}
\SetScale{0.7}
\begin{picture}(300,100)(0,0)
\thicklines
\put(25,110){$H_2$}
\put(25,-10){$H_2$}
\put(265,110){$H_2$}
\put(265,-10){$H_2$}
\put(145,110){$t_L$}
\put(145,-15){$t_L$}
\put(75,50){$t_R^{\,(n)}$}
\put(215,50){$t_R^{\,(m)}$}
\CArc(150,50)(50,-45,45)
\CArc(150,50)(50,45,135)
\CArc(150,50)(50,135,225)
\CArc(150,50)(50,225,315)
\DashLine(50,110)(114.64,85.36){5}  
\DashLine(50,-10)(114.64,14.64){5}
\DashLine(185.36,85.36)(250,110){5} 
\DashLine(185.36,14.64)(250,-10){5}
\end{picture}  
\setlength{\unitlength}{.7pt}
\SetScale{0.7}
\begin{picture}(300,100)(0,0)
\thicklines
\put(25,110){$H_2$}
\put(25,-10){$H_2$}
\put(265,110){$H_2$}
\put(265,-10){$H_2$}
\put(145,110){$\widetilde{t}_L$}
\put(145,-20){$\widetilde{t}_L$}
\put(75,50){$\widetilde{t}_L^{\,(n)}$}
\put(215,50){$\widetilde{t}_L^{\,(m)}$}
\DashCArc(150,50)(50,-45,45){5}
\DashCArc(150,50)(50,45,135){5}
\DashCArc(150,50)(50,135,225){5}
\DashCArc(150,50)(50,225,315){5}
\DashLine(50,110)(114.64,85.36){5}  
\DashLine(50,-10)(114.64,14.64){5}
\DashLine(185.36,85.36)(250,110){5} 
\DashLine(185.36,14.64)(250,-10){5}
\end{picture}  
\vspace{1cm}
\caption{One-loop diagrams contributing to $\lambda_t$.}
\label{figura9}
\end{figure}

The diagrams that contribute to $\lambda_t$ are given in
Fig.~\ref{figura9}, and similar diagrams changing $t\to b$ would
contribute to $\lambda_b$. The result is given by~\cite{topDel}
\begin{eqnarray}\label{radlambdas}
\lambda_{t,b} &=& \frac{3h_{t,b}^4}{8\pi^2}\left\{1-\log(2\pi R M_Z)
+\frac{1}{4(r^2-1)}\left[(r^2-1)\log\left(2-r-\frac{1}{r}\right)
\right.\right.\nonumber\\
&&\left.\left.\phantom{\frac{1^1}{1_1}}+
(1+r^2)\left(Li_2(1/r)-Li_2(r)\right)\right]\right\}
\end{eqnarray}
\vspace{1.cm}
\begin{figure}[H]
\setlength{\unitlength}{.7pt}
\SetScale{0.7}
\begin{picture}(300,100)(0,0)
\thicklines
\put(25,95){$H_2$}
\put(25,5){$H_2$}
\put(265,95){$H_2$}
\put(265,5){$H_2$}
\put(130,110){$\widetilde{t}_R^{(n)},\widetilde{t}_L$}
\put(130,-20){$\widetilde{t}_R^{(m)},\widetilde{t}_L$}
\DashCArc(150,50)(50,0,180){5}
\DashCArc(150,50)(50,180,360){5}
\DashLine(50,95)(100,50){5}  
\DashLine(50,5)(100,50){5}
\DashLine(200,50)(250,95){5} 
\DashLine(200,50)(250,5){5}
\end{picture}  
\setlength{\unitlength}{.7pt}
\SetScale{0.7}
\begin{picture}(300,100)(0,0)
\thicklines
\put(25,110){$H_2$}
\put(25,-10){$H_2$}
\put(265,95){$H_2$}
\put(265,5){$H_2$}
\put(145,110){$\widetilde{t}_L$}
\put(145,-20){$\widetilde{t}_L$}
\put(75,50){$\widetilde{t}_L^{\,(n)}$}
\DashCArc(150,50)(50,0,135){5}
\DashCArc(150,50)(50,135,225){5}
\DashCArc(150,50)(50,225,360){5}
\DashLine(50,110)(114.64,85.36){5}  
\DashLine(50,-10)(114.64,14.64){5}
\DashLine(200,50)(250,95){5} 
\DashLine(200,50)(250,5){5}
\end{picture}  
\end{figure}
\vspace{.5cm}
\noindent where $r=\exp(2\pi i \omega)$.

Minimization of the one-loop effective potential,
$V^\prime_{H_1}=V^\prime_{H_2}=0$ yields the relations
\begin{equation}\label{relaciones}
\tan\beta\simeq \frac{m_t}{m_b},\quad
m_A^2=-m_3^2\,\frac{1+\tan^2\beta}{\tan\beta}
\end{equation}
and the subsequent Higgs mass spectrum is given by
\begin{eqnarray}\label{espectro}
M_H^2 &=& m_A^2\nonumber\\
M_h^2 &=& M_Z^2+4\lambda_t v^2\nonumber\\
M^2_{H^\pm}&=&m_A^2+M_W^2
\end{eqnarray}
The masses, along with $1/R$, turn out to be, as a consequence of the
minimization conditions, functions of $\omega$ and $\mu$, which are
the free parameters of the theory. In particular for $\mu\simlt 1$ TeV
we obtain the bound $1/R\simlt 10$ TeV. This range for values of $1/R$
is in agreement with present experimental bounds on the size of
longitudinal extra dimensions. In particular in five-dimensional
models the strongest bounds are provided by electroweak precision
measurements that yield limits that are model dependent but
generically in the range $1/R\simgt 2-5$ TeV~\cite{bounds}.

What is more interesting, the previous model can be tested by Higgs
searches at the future LHC collider at CERN. In particular LEP
searches in the MSSM Higgs sector settle a lower bound on $m_A$ as
$m_A\simgt 95$ GeV~\cite{LEPbounds}. This bound translates into a
parallel lower bound on $\mu$ which is $\omega$-dependent. The
absolute lower bound turns out to be $\mu\simgt 350$ GeV. On the other
hand the bound $m_A\simgt 95$ GeV also translates into a lower bound
on the Standard Model-like Higgs mass $M_h$ which is again
$\omega$-dependent. The absolute lower limit on the Standard Model
Higgs mass turns out to be $M_h\simgt 145$ GeV. This mass is probably
too large for this Higgs to be discovered at
Tevatron~\cite{HiggsTeV}. In particular if the Standard Model Higgs
turns out to have a mass $\sim 115$ GeV (i.~e. in the final LEP
region~\cite{LEP115}) this model would be excluded. In fact, notice
that for not too large $\mu$ values, $\mu\simlt 600$ GeV,
$M_H<M_h$. However in the large $\tan\beta$ region the state $H$ is
$\sim H_1^0$ with unconventional couplings to gauge bosons and
fermions. In this way the coupling
$HZZ=\left(hZZ\right)^{SM}\cos\beta$ is strongly suppressed as
$\sim1/\tan\beta$ and so is its direct $H$ production at LEP. In short
the Higgs predicted by the above model could not have been discovered
at LEP and its discovery at Tevatron is only very marginal. Its
discovery should only be done at LHC.

\section*{Acknowledgments}
I would like to thank all my collaborators on this subject for so many
fruitful discussions. In particular I.~Antoniadis, K.~Benakli,
M.~Carena, A.~Delgado, S.~Dimopoulos, G.~v.~Gersdorff, N.~Irges,
J.~Lykken, A.~Pomarol, S.~Pokorski, A.~Riotto and C.~Wagner.  This
work was supported in part by CICYT, Spain, under contracts FPA
2001-1806 and FPA 2002-00748, and by EU under contracts
HPRN-CT-2000-00152 and HPRN-CT-2000-00148.




\end{document}